\documentstyle[12pt,aaspp4]{article}

\def\etal{{\it et~al.\ }}
\def\etals{{\it et~al.}'s\ }

\begin{document}
\title{Near-Infrared Galaxy Counts to $J$ and $K$ $\sim$ 24 as a
Function of Image Size\altaffilmark{1}}

\author{Matthew A. Bershady\altaffilmark{2}}
\affil{Department of Astronomy, University of Wisconsin, 475
N. Charter Street, Madison, WI 53706 ({\it mab@astro.wisc.edu})
\\ and \\
Department of Astronomy \& Astrophysics, Pennsylvania State
University, University Park, PA 16802}
\authoremail{mab@astro.wisc.edu}

\author{James D. Lowenthal\altaffilmark{2}}
\affil{Department of Physics and Astronomy, University
of Massachusetts, Amherst, MA 01003 ({\it james@velo.phast.umass.edu}) }
\authoremail{james@velo.phast.umass.edu}

\author{and}

\author{David C. Koo}
\affil{University of California Observatories / Lick Observatory, 
Department of Astronomy and Astrophysics, University of
California, Santa Cruz, CA 95064 ({\it koo@ucolick.org}) }
\authoremail{koo@ucolick.org}

\altaffiltext{1}{Based on observations obtained at the W. M. Keck
Observatory, which is operated jointly by the University of California
and the California Institute of Technology.}

\altaffiltext{2}{Previous affiliation: Hubble Fellow, University of California
Observatories / Lick Observatory}

\begin{abstract}

We have used the Keck 10m telescope to count objects as a function of
image size in two high galactic latitude fields covering 1.5
arcmin$^2$ and reaching 50\% completeness depths of $K$=24 and
$J$=24.5 for stellar sources. Our counts extend $\sim$1 magnitude
deeper in $K$ than those of surveys with other telescopes; complement
other Keck surveys in the $K$ band that provide counts at comparable
or shallower depths but that have not utilized image structure; and
extend by several magnitudes the $J$ band counts from brighter surveys
using smaller telescopes that cover larger areas. We find the
surface-density of objects at $K$=23 to be higher than previously
found ($\sim$500,000 mag$^{-1}$ deg$^{-2}$), but at $K<22$ to be
consistent with other surveys. The slope of the $K$ band counts ($d
{\rm log} A / dm = 0.36$) is similar to others near this depth as well
as to our own $J$ band counts (0.35). Counts in $J$ and $K$ bands are
both in excess of our empirical no-evolution models for an open
universe, with the largest excess observed in $J$. The counts are a
factor of 2 higher than mild-evolution models at $J$ and $K$ $\sim$
23. The slope of the model counts is {\it insensitive to the assumed
geometry} even in the near-infrared primarily because the {\it model}
counts are dominated by low-luminosity ($< 0.1$ L$^*$) objects at
modest redshift ($z<1$) with small apparent sizes (half-light radii
$\leq 0.4$ arcsec, i.e. $<4$ h$^{-1}_{50}$ kpc). The observed counts
rise most steeply for these smaller objects, which dominate the counts
fainter than $K$=22.3 and $J$=23.3. However, {\it the greatest excess
relative to no-evolution models occurs for the apparently larger
objects} which have a median $J-K$ of $\sim 1.5$. At these depths, the
size and colors of such objects correspond equally well to luminous
($\geq0.1$ L$^*$), {\it blue} galaxies at $1<z<4$, or progressively
more diffuse, {\it blue}, low-luminosity (0.001-0.1 L$^*$) galaxies at
$z<1$.  The majority of these sources are too faint for spectroscopic
measurement. Based on optical colors, we can rule out that the excess
is due to very low luminosity ($< 0.0001$ L$^*$) {\it red} galaxies at
$z<0.25$. We also find a {\it deficit} of galaxies with red $J-K$
colors corresponding to non-evolving, luminous, early-type (i.e. ``red
envelope'') galaxies at $1<z<3$. Even assuming the deficit is due to
their appearance as blue galaxies, they could account only for 10-30\%
of the excess of large, blue galaxies. The nature and redshift
distribution of excess large and small galaxy populations at $K$=24
and $J$=24.5 remain indeterminate from these data alone.


\end{abstract}


\clearpage
\section{Introduction}

Near-infrared galaxy counts are considered by some a panacea for the
study of galaxy evolution and cosmology. One reason is that compared
to counts at optical wavelengths, near-infrared counts are less
sensitive to the presently unknown amounts of dust reddening within
galaxies, as well as to the uncertain evolution of stellar populations
with look-back time. Moreover, in the near-infrared, $k$-corrections
are well known to high redshift because they are based on well studied
optical photometry of local galaxies (e.g. Bershady 1995). Yet the
flatness of the $K$ band counts relative to optical counts has been a
puzzle which has so far eluded satisfactory -- or at least an
agreed-upon -- explanation (e.g. Cowie \etal 1990, Gardner \etal 1993,
Gronwall \& Koo 1995, Djorgovski \etal 1995, Yoshii \& Peterson
1995). This is due to the difficulty of interpreting galaxy counts
(e.g. Koo 1990, and references therein), which represent a convolution
of a detection function, the multi-variate distribution of rest-frame
galaxy properties (such as luminosity, color and size), and the
cosmological volume. Thus counts alone -- even in the near-infrared --
are unlikely to provide sufficient information to disentangle these
various components and discern the true nature of objects at the
faintest limits of observation.

The counts are, however, generally derived from measurements of
images, which have the potential to yield additional data.  Colors,
for example, have been used to infer redshifts beyond the reach of
spectroscopy (e.g. Lilly \etal 1991). Another approach is to exploit
image structure at the depths reachable with 4-10m class telescopes
(Lilly \etal 1991, Colless \etal 1994), but these efforts are still in
their infancy. With excellent seeing of 0.3-0.6 arcsec routinely
obtainable in the near-infrared from the ground, object size,
surface-brightness, concentration, and asymmetry can all be measured
and added to counts, magnitudes, and colors for constraining models to
the faintest limits reachable from the largest telescopes. For
example, between redshifts of $\sim$0.8 and 3.5, then there is little
change in apparent size for a given metric size (e.g. $<$ 20\% change
for 0.1$<$q$_0$$<$0.5). If, in addition, the tight correlation between
metric size and luminosity already observed for bright samples of
nearby galaxies persists to higher redshift, apparent size and
luminosity should be well correlated. In other words, {\it apparent
size can be used to estimate luminosity.}

There are caveats: A significant space-density of low
surface-brightness galaxies, if they exist, will broaden the observed
size - luminosity relation, particularly if they are more readily
detected in deep images (e.g. Ferguson \& McGaugh 1995).  Evolution
via merging may also alter this relation, as may any wavelength
dependence of the observed galaxy size.  Nonetheless, faint surveys
sample the moderate-redshift universe at very low luminosities
inaccessible to brighter surveys. Therefore it is worthwhile to assess
whether the faintest sources are predominantly large or small objects.
Naively, one might expect the small objects may represent a class of
low luminosity objects at low to moderate redshifts, missed by
brighter surveys, while the larger objects may be luminous galaxies at
high redshift.  In contrast, recent evidence appears to indicate that
luminous (L$^*$), high redshift ($z>3$) galaxies are predominantly
small (Giavalisco \etal 1996, Lowenthal \etal 1997). These
measurements, however, have been made in the rest-frame ultra-violet
for a select sample of galaxies; the near-infrared image structure for
complete, magnitude-limited samples is not known.

In this paper, we consider these extra dimensions of information by
presenting very faint, near-infrared counts in the $J$ and $K$ bands
as a function of image size from images taken with the Keck 10m
telescope. We have taken advantage of the excellent seeing ($\sim$0.6
arcsec FWHM), large telescope aperture, and low backgrounds to obtain
very deep, but small ($\sim$0.7 arcmin$^2$) images of two
high-galactic latitude fields. To assess the nature of the galaxies in
our sample, we construct a null hypothesis for the counts and colors
as a function of image size at faint magnitudes by adopting a
no-evolution model that is based on observations of brighter,
multi-color sample (Bershady \etal 1994). We then determine how the
observed counts and size and color distributions deviate from this
well-defined prediction. This approach has the advantage that it
obviates all but cosmological model assumptions. Hence the measured
differences between model predictions and observations can be
attributed cleanly to failings of the model or to galaxy evolution
without recourse to fine-tuning model-specific parameters.

The organization of the paper is as follows: Section 2 details our
field selection and observations, followed by a description of the
image processing (Section 3), object detection (Section 4), and
photometry and measurements of counts (Section 5). A data table of
sizes and magnitudes for individual sources is presented therein. Our
analysis of the observed $J$ and $K$ band counts and model predictions
are presented in Section 6, while the sizes, colors and evolution are
discussed in Section 7. The main findings of our survey are
summarized in the concluding Section 8. An appendix contains a
description of how counts are derived from images of non-uniform depth

\section{Field Selection and Observations}

Several fields where multi-band optical photometry already existed at
high galactic latitudes were chosen for near-infrared imaging. Our two
deepest near-infrared fields (Figures 1 and 2 [plates ???])  are SA~57
6575 (part of an area studied by Hall \& MacKay [1984] using deep CCD
drift scans in two bands similar to $R$ and $I$, hereafter SA~57), and
Herc-1 5677 (part of an area where deep $U$ CCD data have been
obtained by Majewski, Koo \& Kron, hereafter Herc-1). Our selection
criteria for these fields were (i) that they be centered on a star
sufficiently bright to provide accurate registration and point spread
function (PSF) determination, but not too bright to cause problems
with scattered light, and (ii) that the apparent surface density of
faint galaxies as visually assessed in the deep optical CCD data
($R\sim 25$) appeared representative of a much larger region covered
in the optical images.  We intentionally did not select `blank'
fields, since such a choice would bias a survey to sampling
systematically under-dense lines of sight. This bias could be
substantial even at faint magnitudes if low luminosity galaxies are
inherently spatially correlated with more luminous galaxies, which
have been `avoided' by selection of `blank' fields.

Observations were conducted in $J, H, K, K',$ and $K_s$
bands\footnote{Most `standard' $K$ bands cover between $\sim$2-2.4
$\mu$m (e.g. Bessell \& Brett 1988), whereas the $K'$ band (Wainscoat
\& Cowie 1992) covers between $\sim$1.9-2.3 $\mu$m and the $K_s$ (`K
short') band covers $\sim$2-2.3 $\mu$m. $K'$ and $K_s$ are designed to
avoid the rising thermal background at the red end of the $K$
window. The possible advantage of $K'$ over $K_s$ is greater
bandwidth, but this depend critically on the transparency (and
emissivity) of the atmosphere at the blue end of the $K$ window. The
specific NIRC filter half-power points at $77^{\circ}$ K are:
1.105-1.397 $\mu$m ($J$), 1.491-1.824 $\mu$m ($H$), 2.000-2.427 $\mu$m
($K$), 1.955-2.292 $\mu$m ($K'$), 1.99-2.32 $\mu$m ($K_s$).} on three
nights (April 24-27, 1994) using the Keck I telescope and the NIRC
camera\footnote{256$^2$ SBRC InSb array, 0.15 arcsec/pixel (Matthews
\etal 1994).}. Conditions were almost totally photometric, with seeing
ranging from 0.4-1.5 arcsec FWHM, with a median of 0.6 arcsec. Target
exposures were 10 sec in length, with a dwell time of 90 sec per
position. A dither sequence of 11 pointings was used, with
characteristic offsets of 9 arcsec; we also adopted additional random
offsets of 1 to 3 arcsecs between sequences. Typical backgrounds in
$J$ and $K'$ were 15.7$\pm$0.1 and 13.5$\pm$0.1 mag arcsec$^{-2}$
respectively.  Six faint standard stars\footnote{FS 15, 21, 20, 19,
27, and 33, (Casali \& Hawarden, 1992)}, covering a range of
$-0.22<J-K<0.42$ and $-0.14<H-K<0.06$, were observed in $J, H, K, K'$,
and $K_s$ bands at several exposure times. While most of our target
data was observed in $J$ and $K'$ bands\footnote{$K'$ yielded the
highest signal to noise (S/N) compared to $K$ or $K_s$, which had
backgrounds of 13.6$^{+0.1}_{-0.35}$ and 13.75$\pm$0.1 mag
arcsec$^{-2}$ respectively.}, the standard stars and target frames
taken in the $K$ band were used to calibrate directly to the standard
$K$ band. Magnitudes and colors are quoted here in the Vega system.
We estimate photometric zero-point uncertainties to be $\leq$2\%, but
we note that all of the standards were substantially bluer than most
of the target sources.

\section{Image Processing}

Prior to image reduction, a series of images of the warm dome covering
a large range of exposure times (and hence detected count levels) were
analyzed to determine the linearity of the NIRC camera response. The
resulting curve of count rate vs. total counts indicated that, for all
applications in this program, the departure from linearity would 
be less than 1.5\% and thus small enough to ignore.

All images were first bias subtracted using ``master'' bias frames;
such frames were constructed from clipped averages of multiple dark
frames with minimum integration times (0.43 sec). Flat field
calibration images were then constructed for each band on each night
by combining (with a median filter) all the unregistered data obtained
in that band on that night, and these were used to flatten all the
individual data frames. A map of bad (i.e., hot or dead) pixels was
assembled by examining the pixel statistics of a large number of
images and rejecting those pixels that either varied widely from frame
to frame or stood consistently more than $10\sigma$ above or below the
mean. A sky image, made from a median of the 9 frames closest in time
and with the same filter, was subtracted from each frame.

A preliminary image stack was then used to make a mask image that
indicates the pixel positions brighter than $2\sigma$ above the sky
background. Typically 1\%-2\% of the total pixels in the image were
included in the mask image. The sky subtraction was repeated on each
original frame, using the mask to exclude bright objects from each
individual sky image before building the composite sky frame. Bad
pixels and pixels previously flagged as affected by cosmic rays were
also excluded. This initial data reduction used the software package
DIMSUM within IRAF (Stanford \etal 1995).

The flatness of the individual processed images was found to be few$
\times10^{-5}$ of sky on scales of 15 to 35 arcsec. To remove these
small remaining gradients, a second sky-subtraction was performed
using a first-order cubic spline fit to the unmasked pixels of each
image. A final image stack for each field and band was made by first
registering all images via cross-correlation of the brightest 6-10
objects, and then weighting each image by the S/N of the bright
central star as determined from multi-aperture photometry in each
image. S/N was defined to be proportional to the ratio of the
photon counts within the half-light radius to the effective sky-noise
within the half-light radius (close to the maximum S/N). In this way
the S/N explicitly took into account changes in seeing, transparency,
and background. The weighting process substantially improved the image
depth and quality, with the final flatness over large scales being
$<3\times10^{-6}$ of sky. For object detection with FOCAS (but not for
photometry; see below), the final image stacks were scaled by the
square-root of the exposure maps to normalize the noise to a constant
value across the image (as displayed in Figures 1 and 2).

\section{Object Detection}

There is always a trade-off between completeness, which we define as
the depth where 50\% of the objects are detected) and reliability
(percentage of real versus spurious detections) at that depth. If the
reliability function can be accurately determined, object detection
can be pushed to fainter limits by relaxing the detection criteria to
improve completeness. In practice, reliability is difficult to measure
unless one has additional (and preferably deeper) data to make an
independent assessment. Since we are pushing to the very faintest
limits at high backgrounds, unknown sources of electronic noise and
known (but poorly characterized) imperfect camera baffling both led
use to be cautious and choose detection parameters that minimized
spurious detections.

Completeness and reliability also depend on the image structure of the
objects. Smaller or more concentrated (higher surface-brightness)
objects are more readily detected at a given total magnitude, but
spurious detections tend to be more frequent. In addition to being
interested in counting sources as a function of image size, we are
compelled to do so simply in order to estimate correctly the total
number of sources.

We have used FOCAS (Jarvis \& Tyson 1981, Valdes 1982) to detect
sources. We have also tested the completeness and reliability as a
function of the FOCAS detection parameters: minimum area, isophotal
threshold, and detection kernel. For a grid in this parameter space,
we determined detection completeness as a function of total magnitude
and image size (defined below), using $\sim$10$^4$ simulations for
each of the four final, stacked images.  Test sources consisted of the
brightest 6 objects in both fields ($K$$<$19) that spanned the
observed range of image size, including the PSF.  These templates were
artificially dimmed and added back into the original images at random
locations. FOCAS was then run with the same detection parameters to
search for the simulated objects. As expected, the completeness scaled
with the square-root of the effective exposure time.

Detection reliability was measured by photometering all detected
objects on a pair of images constructed from each of two randomly
chosen but exclusive halves of the data (for each field and band) in
circular apertures equivalent to the isophotal area as detected in the
full image stacks. Objects with magnitudes differing by more than 5
sigma (as determined by the sky noise and photometry aperture) were
considered to be spurious detections. When the detection parameters
yielded few such deviant points, the distribution of magnitude
differences appeared bimodal, and visual inspection of the
``spurious'' detections nominally confirmed the numerical result. The
reliability in this case is a steeply dropping function of flux,
beginning near the 50\% completeness limit. With detection parameters
yielding larger numbers of deviant points, the bimodality disappeared;
similarly it became difficult to assess detection reliability
visually. Such detection parameters were excluded from further
consideration. It is worth noting that additional simulations showed
that reliability would be substantially {\it overestimated} by
counting sources detected in simulated, blank, Gaussian random noise
fields. This method was adopted by Djorgovski \etal (1995).

Based on the above tests, we chose a +3/-6 sigma isophotal threshold,
a minimum area corresponding to that within the FWHM of the PSF in
each image, and the PSF as the detection kernel. This set is optimal
for detecting unresolved sources.  In addition to the quantitative
estimates of completeness and reliability, we visually inspected the
images to check that the chosen parameters appeared to
detect all apparently real objects, while minimizing the
number of spurious detections.  The isophotal threshold corresponds to
a surface brightness in the kernel-convolved images of 24.25, 25.5,
23.9, 25.0 mag-arcsec$^{-2}$ for the deepest regions of SA~57 $K$ and
$J$, and Herc-1 $K$ and $J$, respectively. These numbers are
indicative of our ability to detect large, low surface-brightness
objects. More relevant for the detection of compact sources are the
magnitudes corresponding to the 3$\sigma$ detection limits for flux
within the minimum detection area. These correspond to 25.1, 25.8,
24.5, and 25.5 mag, again for the deepest regions of SA~57 $K$ and
$J$, and Herc-1 $K$ and $J$, respectively. The S/N at 50\% detection
limits, however, is around 5, independent of image size.

\placefigure{fig3}

Figure 3 illustrates the completeness as a function of $K$ magnitude
and image size for the deepest portion of the SA~57 6575 field. Table 1
lists, in brackets below each heading, (i) the 50\% detection limits
for each object image size; and (ii) the reliability at these
limits. Note that the difference in depth between fields depends on
image class because of changes in seeing; seeing is worse in Herc-1
for the $K$ band and worse in SA~57 for the $J$ band. Also note the
{\it large} ($\sim$0.5 mag) difference between the 50\% detection
limits as a function of image size for a given field and band. Stellar
sources have 50\% detection limits $\sim$0.5 mag fainter than either
of our two categories, $s$ and $l$, defined below. {\it Because
completeness falls rapidly with magnitude, accounting for such
differences is essential to provide reliable corrected source counts
near the detection limits.}

\section{Photometry, Sizes and Counts}

Final photometry consists of 1.8-2.1 arcsec diameter, fixed-aperture
magnitudes corrected to 'total' on the basis of object size. The
choice of aperture varied from image to image, according to the
seeing, in order to make the aperture corrections at most -0.35 mag
(as determined empirically from brighter sources in each image and by
photometering artificial objects of known size, shape, and brightness
that were added into the real images). Object size is defined by the
$\eta$-function (Petrosian 1976), using the convention of Kron (1995)
where $\eta$ is the ratio of the surface brightness at radius $\theta$
to the average surface brightness interior to $\theta$. Sizes were
measured from logarithmically spaced, multi-aperture photometry for
values of $\eta$=0.5 and 0.1 using the algorithm described in Wirth
and Bershady (1998). Such sizes depend only on the surface-brightness
distribution and not amplitude, and hence are metric radii.\footnote{A
'metric' radius is defined to mean a measure of size corresponding to
the same physical scale for galaxies of the same physical size and
light distribution.} The uncertainty in the measured apparent sizes
$\theta_{\eta}$ is about 50-75\% at the detection limit for $\eta$=0.5,
and somewhat worse for $\eta$=0.1 (in general
$\theta_{0.5}<\theta_{0.1}$). 'Total' magnitudes are defined as the
light enclosed within the $\eta$=0.1 radius. With zero-points set from
stellar sources using the same magnitude scheme (e.g. see Bershady
\etal 1994), and in the absence of noise, these 'total' magnitudes are
within +0.00, +0.02 and +0.10 mag of the true total value for
Gaussian, exponential and r$^{1/4}$-law profiles,
respectively. Respectively, $\theta_{0.1}$ radii are equivalent to
2.28, 2.86 and 4.39 times these profiles' half-light radii.

For the purpose of counting, we have defined two bins in apparent size
$\theta_{0.5}$, adjusted to give the same true apparent size (in the
absence of image blur) on each image. Adjustments were made on the
basis of simulations of exponential profiles (at various inclinations)
and r$^{1/4}$-law profiles (with a range of ellipticities), which all
yielded very similar changes in $\theta_{0.5}$ as a function of seeing
and intrinsic half-light radius. The specific size ($\theta_{\eta}$)
that divides the two image classes (small, $s$, and large, $l$) are:
$\theta_{0.5}$ = 0.62, 0.81, 0.71 arcsec respectively for SA~57 $K$,
$J$, and Herc-1 ($J$ and $K$). The simulations indicated that the {\it
intrinsic} size is $\theta_{0.5}$ $\sim$ 0.44 arcsec, i.e. in the
absence of image blur. This size corresponds to 3.75 kpc at $z = 1$,
or exponential disk scale lengths of 2.1 h$_{50}^{-1}$ kpc (h$_{50}$ =
H$_0$ / 50 km s$^{-1}$ Mpc$^{-1}$, q$_0$ = 0.5). For Gaussian,
exponential and r$^{1/4}$-law profiles, $\theta_{0.5}$ radii are
equivalent to 1.35, 1.07, and 0.17 times these profiles' half-light
radii respectively (1.6 $\sigma$ for a Gaussian, 1.8 scale-lengths for
an exponential).

The raw and corrected counts are listed in Table 1 for each field,
band, and image size. Corrections take into account completeness,
reliability, and the usable area as a function of depth in
half-magnitude intervals. (The images are of non-uniform depth because
of dithering; the Appendix describes how the counts are constructed
using the full area.) The values in Table 1 have been averaged over
one magnitude intervals, but are listed every 0.5 mag, and hence
adjacent bins are correlated. Errors include counting statistics
(Gehrels 1986) added in quadrature to estimated uncertainties in the
completeness corrections based on the variance in the simulations for
the set of templates for each object class (typically $\sim$5\%). {\it
The counts for each band in a given field are entirely independent.}
The total numbers of sources represented in Table 1 are 163 for the
$K$ band and 118 for the $J$ band.  This sample size is somewhat
greater than for other near-infrared surveys of similar depth,
e.g. 111 and 88 sources to $K\sim24$ by Djorgovski \etal (1995) and
Hogg \etal (1997), respectively.

A catalogue of $JK$ magnitudes, sizes, and positions for the
individual sources used to produce the counts in Table 1 are listed in
Table 2, in two-column format. This table also includes several bright
sources in bins brighter than listed in Table 1. Sources are sorted by
$K$ magnitude (or $J$ magnitude, if only detected in $J$) seperately
for each field. For sources detected in only one band, upper limits
are tabulated for the other band where possible. Such sources are
identified by the absence of photometric error estimates, as described
in the table notes. Other details of the data set are noted therein,
including astrometric formulae for transforming the relative x,y pixel
locations into Righ Ascension and Declination. We estimate that the
relative and absolute astrometry is accurate to $\sim$0.3 arcsec or
better.

While Table 2 identifies sources as either small ($s$) or large ($l$),
note we have not distinguished stars from galaxies for the following
reasons: (1) There is no clear stellar locus in size-magnitude
diagrams, compared to e.g. Kron (1980, Figure 9). (2) On the basis of
simulations, we found FOCAS correctly classified stars and galaxies
only 50\% of the time by $K$=22.5 for our deepest field. (3) The
apparent lack of stars may be real. Other studies at high galactic
latitude find 10\% contamination at $K$=19.5-20, consistent with
models that predict the fractions of stars to drop to 2\% at $K$=21.5
(Cowie \etal 1994, McLeod \etal 1995). Any stars will be small ($s$)
objects in our sample.  Most of our detected $s$ sources are redder
than $J-K\sim1.25$, which is redder than cool giant and main sequence
stars. Eight are bluer than this value, but are coincident in color
and magnitude with some $l$ sources. While there remains the
possibility that we have detected some extremely faint, red Galactic
stars, it is more likely that these are compact galaxies. Two-color
photometry is one way to resolve this issue in the future.

\section{$J$ and $K$ Band Galaxy Counts}

\subsection{Discrepancies Between Surveys?}

\placefigure{fig4}

What can we infer about faint galaxies from their sizes, colors, and
number?  We begin by comparing our counts, summed over sizes and
averaged over fields, to counts from other surveys as well as models
(see Figure 4). The last magnitude interval where all image sizes have
detection completeness above 50\% in our survey is $K$=23 (SA~57 only)
and $J$=23.5 (both fields). In order not to introduce an artificial
jump in the last $K$ magnitude bin due to field-to-field variations
between SA~57 and Herc-1, we have plotted the SA~57 counts in the
faintest bin scaled according to their ratio to the average counts,
averaged over the previous two bins. For most other surveys no changes
have been made to their photometry since their schemes are either
comparable to ours or insufficiently specified to attempt
adjustment.\footnote{A recent paper by Hogg \etal (1997) contains
galaxy photometry to $K<24$, but provides insufficient data to
estimate counts. In particular, while they state their detection is
90\% complete to $K=23$, we estimate based on raw counts constructed
from their tabulated source list that there is substantial
incompleteness beyond $K=22.5$ over their full survey area. For this
reason we have not considered their data for counts of galaxies.}
However, Djorgovski \etal (1995) employed aperture corrections for
objects fainter than $K$$=$21 that assumed a stellar curve of
growth. We find that this assumption results in underestimating the
true flux of galaxies and over-estimating the depth of their survey by
0.5 mag. Therefore we have applied magnitude corrections to their
faint counts to make them consistent with the average aperture
corrections we have applied to our data.

Our counts are consistently higher than those of Cowie \etal (1994)
and Djorgovski \etal (1995), but are bracketed in amplitude and slope
by those of Soifer \etal (1994) and McLeod \etal (1995). (The McLeod
\etal survey includes two fields within $\sim$20 arcminutes of our
SA~57 field, and one field within 9 arcmin of our Herc-1 field.)
While our counts agree roughly with the results of Moustakas \etal
(1997) for $K<21$, our counts are in excess of their counts at fainter
magnitudes (a regime where our data are significantly more
complete). Since the Soifer \etal (1994) and McLeod \etal (1995)
samples end at $K$$\sim$21.5, our counts represent a substantial
increase in the number of measured galaxies at $K$$\geq$21.5. The
slope of our $K$ counts (dlog($A(K)$)/dm$ =
0.36\pm0.02$)\footnote{Slope uncertainties quoted throughout are 67\%
confidence intervals for one free parameter.} is slightly shallower
than Djorgovski \etals (1995) as plotted in Figure 4. Had we not made
the large $\sim 0.5$ magnitude corrections to Djorgovski \etals (1995)
counts, we would then have comparable slopes but more discrepant
amplitudes. With the exception of Moustakas \etal (1997) who report a
slope of $\sim$0.23, counts from all other surveys are substantially
steeper than the value of 0.26 reported by Gardner \etal (1993) which
was based on data from Cowie \etal (1994).  Indeed, both this survey
and that of Djorgovski \etal (1995) yield count slopes which do {\it
not} decrease for $K$$>$21.5, but show some hint of an increase, in
contrast to the Gardner \etal and Cowie \etal results.  Our $J$ band
counts also show no sign of a flattening slope at the faint end, and
reach surface densities equivalent to our values at $K$$\sim$22.7, but
well in excess of values from other faint $K$ surveys at this
depth. The $J$ counts have comparable slope to our $K$ band counts
(dlog($A(J)$)/dm=0.35$\pm$0.04). There are no $J$ data in the
literature for an independent direct comparison.

To what extent are the variations in counts and slopes due to large
scale structure? Within our own data, field-to-field variations in the
counts are within the Poisson counting noise for a given magnitude bin
but do vary systematically with magnitude. Herc-1 is 20\% higher than
SA~57 for 19.5$\leq$$K$$\leq$20.5, while SA~57 is 30\% higher than
Herc-1 for 21$\le$$K$$<$22.5. A similar trend occurs in the $J$
counts, with Herc-1 having a higher surface-density at brighter
magnitudes (21$\leq$$J$$\leq$22). Slopes for the individual fields are
0.35$\pm$0.02, 0.27$\pm$0.05 for 19.5$<$$K$$<$22.5 and 0.46$\pm$0.04,
0.34$\pm$0.06 for 20.5$<$$J$$<$23.5 for SA~57 and Herc-1 respectively.
These $K$ band slopes are within -- but span-- the bounds observed by
others.

Is either of our fields representative? We have addressed this
question using a cell-count analysis of the deep $KG3$ ($\sim$$R$)
band catalogue covering 30.2 arcmin$^2$ from the drift-scan survey of
Hall \& Mackay (1984).  We find that our SA~57 field is low relative
to the average cell of its size by about 30$\pm$17\% over the
magnitude range 20$<$$KG3$$<$25, but becomes more representative at
the faintest magnitudes, yet still low by 15$\pm$10\%.
Surface-densities at $KG3=25$ are comparable to what we find at
$J$$\sim$21.5 and $K$$\sim$20.5. For Herc-1, we have checked against a
photographic catalogue to $R_F$=23 covering 0.384 deg$^2$ (Munn \etal
1997) to find that for 21$<$$R_F$$<$23, Herc-1 has a 30$\pm$34\%
surfeit of galaxies. The surface-density of objects to $R_F$=23 (Kron
1980) is comparable to the surface-density at $K$=19. While these
checks are rather uncertain, they are consistent with Herc-1 being
somewhat unrepresentatively high at the bright end of our
near-infrared counts. We have, however, no independent optical check
at this time for the faint end of our near-infrared counts.

Given the small size and great depths of these fields, it is plausible
that large-scale structure is producing variations in the counts as a
function of magnitude. Over all magnitudes, the field-to-field
variations within our sample are at a $\sim$30\% level, two-thirds of
which is expected from Poisson noise. The remainder is only slightly
higher than the $\sim$10\% variation expected from clustering in
images of this size and depth estimated by Djorgovski \etal
(1995). The consistently lower counts of Djorgovski \etal (1995) and
Cowie \etal (1994) are also plausibly due to real variations in the
counts. Larger, ultra-deep surveys are needed to resolve this issue.

\subsection{A comparison to models}

\placefigure{fig5}

\subsubsection{The models of Gronwall \& Koo}

It is instructive to compare the current data to models, two sets of
which are shown in Figure 4. The models of Gronwall \& Koo (1995)
predict that the counts do {\it not} roll over at $K$$\sim$22, and may
in fact steepen beyond $K$=23, much like our data. These models
include both passive evolution and the effects of internal extinction
due to dust, both of which are included in the derivation of
luminosity functions. The luminosity functions are constrained to fit
observational data of faint galaxies. These data include counts and
distributions in color and redshift for a wide range of surveys, but
they do not include our, Djorgovski \etals (1995), or Moustakas \etals
(1997) $K$ band counts. Our counts are in excess of those predicted by
the Gronwall \& Koo model for $K>21.5$ and for all magnitudes in the
$J$ band well constrained by these data ($21.5<J<23.5$).  The excess
reaches about a factor of 2$\times$ for $K=23$ and $2.7\times$ for
$J=23.5$. For log($A$) $\sim$ 5.2 mag$^{-1}$ deg$^{-2}$, the excess is
$\sim$ 40\% in $K$ and a factor of two in $J$. This occurs at $K=22$
and $J=23$, implying a mean galaxy color contributing to this excess
of $J-K=1$. This is about 0.5 mag bluer than the median color at these
magnitudes, but objects with these colors are plentiful in our sample,
as we will show. Figure 5 shows that based on $J-K$ color, such
galaxies would lie at very low redshift ($0.05<z<0.25$) if there is no
color evolution. Alternatively, these galaxies could have a much wider
range of possible redshifts if they are undergoing a very strong burst
of star formation seen very early on in the burst (also illustrated in
Figure 5). Hence, the usual ambiguity between strong evolution in
color and luminosity and a non-evolving, steep faint-end luminosity
function applies to the interpretation of these data (e.g. Koo \&
Kron, 1992).

\subsubsection{1/V$_{max}$ empirical simulations}

The second set of models, labeled 1/V$_{max}$, are empirical
simulations. They are based on a $U$- though $K$-band observational
survey of low redshift galaxies (Bershady \etal 1994), which has been
scaled, object by object, by the relative accessible volumes in the
input survey and the output simulations. Hence galaxy evolution does
not enter into these models, and the only free parameters are those
that specify the curvature (q$_0$ and $\Lambda_0$; we assume
$\Lambda_0$=0 here). Because the input survey has $U$ through $K$ band
photometry, $k$-corrections in the $K$ band are determined empirically
for each simulated object\footnote{This is true for $z\leq$7, which is
not exceeded in our simulations.} (Bershady 1995). As a result, the
effects of internal extinction due to dust are empirically accounted
for; this is not a model parameter. However, the input sample, while
it extends 7.5 mag fainter than $M^*_K$ (0.001 L$^*_K$), contains few
galaxies (16) fainter than $M^*_K$+5 (0.01 L$^*_K$) because it is a
relatively small, magnitude-limited sample (roughly $B$$<$20.5 and 0.9
deg$^2$). Hence estimates of the contribution from dwarf galaxies
remain somewhat uncertain. 1/V$_{max}$ models substantially
underpredict our observed counts by $K=20$, but stay within a factor
of 2 until $K>22$ for q$_0$=0.

\subsubsection{The contribution of low luminosity galaxies to the counts}

Can the models' underprediction of the observed counts plausibly be
explained by low luminosity galaxies, e.g., missing in the 1/V$_{max}$
input sample? Low luminosity galaxies are expected to be present in
significant numbers by $K \sim 20$, but the precise contribution
depends on the currently unknown faint-end of the local $K$ band
luminosity function. Remarkably, a 0.01 L$^*_K$ galaxy can be seen to
$z \sim 0.8$ at $K = 23$, but at this distance only 3-10\% of the
survey-limited volume is included. This estimate depends on q$_0$ and
assumes a redshift upper limit of $z=4$, where $K=23$ corresponds to
L$^*_K$. Nonetheless, galaxies in our 1/V$_{max}$ models with $0.001 <
$ (L/L$^*$)$_K < 0.01$ contribute roughly 30\% and 60\% of the counts
at $K = 23$ for q$_0$ = 0 and 0.5, respectively.  

Lower-luminosity galaxies, however, are expected to contribute
insignificantly to the faint counts: The slope of the $K$ band
luminosity function for the 1/V$_{max}$ models' input survey (Bershady
\etal 1998) is well described by $\alpha = -1.6$ for $M_K < -23$
($\sim 0.06$ L$^*_K$, H$_0$=50, q$_0$=0). Using this slope to
extrapolate the observed space densities fainter than 0.001 L$^*_K$,
we find that galaxies with 0.0001 $<$ (L/L$^*$)$_K < 0.001$ would
contribute only 5\% and 12\% of the counts at $K = 23$ for q$_0 = 0$
and 0.5, respectively. Therefore, the largest uncertainties in the
predicted 1/V$_{max}$ model counts at the depths probed in our study
here come from the luminosity function in the range 0.001 $<$
(L/L$^*$)$_K < 0.01$, {\it unless} there is a very strong upturn at
even fainter luminosities in the already steep observed slope of the
luminosity function.

Consider, for example, how much the faint-end luminosity function
slope must be steepened in the range 0.001-0.01 L$^*$ to fit the
counts at $K=23$ with little or no evolution.  Referring to Figure 4,
this requires $\sim 15 \times$ increase in the integrated volume
density of such galaxies for the 1/V$_{max}$ models (q$_0$=0 and 0.5),
and between a factor of 6.5 and 8.5 increase for the models of
Gronwall \& Koo (no evolution and mild evolution, q$_0$=0.05). This
corresponds roughly to changing the faint end slope $\alpha$ (as
parametrized by the Schecter function) by $\Delta\alpha \sim -0.4$
over the same modest luminosity range for the 1/V$_{max}$ models, and
less over larger ranges of luminosity. For the models of Gronwall \&
Koo, $\Delta\alpha \sim -0.3$ would suffice. Such steeper slopes
cannot be ruled out, since recent results from local surveys yield
values of $\alpha$ discrepant by at least this amount (e.g. compare
results from Marzke \etal [1994] with those of Loveday \etal [1992] or
Lin \etal [1996]). The volume density of such low-luminosity galaxies
is thus not well constrained at any redshift. Certainly $\Delta\alpha
\sim -0.3$ would be possible within the uncertainties of the
luminosity function determined from our $B<20.5$ sample or from the
deeper surveys of Lilly \etal (1995) and Ellis \etal (1996). With such
uncertainties, the amount of evolution needed to fit the counts
remains unknown. If q$_0$=0.5, either substantially more evolution or
an even steeper luminosity function is needed than in the low q$_0$
case. We emphasize that this conclusion is based not on the {\it
slope} of the counts, but rather on their {\it amplitude.}

\subsubsection{Cosmological interpretation of the count slope and amplitude}

\placefigure{fig6}

The 1/V$_{max}$ models and those of Gronwall \& Koo allow us to
explore the sensitivity of the counts to luminosity function
parameters and, in turn, whether the counts can be used reliably to
probe q$_0$. While the 1/V$_{max}$ models substantially underpredict
our observed counts, they only moderately underpredict the count slope
for $K>20$. For q$_0=0$, they yield close to the same {\it slope} as
the models of Gronwall \& Koo (1995). In the $J$ band, the 1/V$_{max}$
models predict even fewer galaxies than either the data or the models
of Gronwall \& Koo (1995), but again the {\it slopes} are
comparable. Unlike other no-evolution models (e.g. Fukugita \etal
1990, as used by Gardner \etal 1993, or Yoshii \& Takahara 1988, as
used by Cowie \etal 1990 and Djorgovski \etal 1995), for q$_0=0.5$,
our 1/V$_{max}$ model predicts that the counts do {\it not} roll over
around $K=22$, but instead continue to rise beyond $K=25$ and
$J=25$. This reflects a steeper faint-end slope of the local $K$ band
luminosity function than previously adopted by model builders. Yoshii
\& Takahara (1988) and Fukugita \etal (1990) both adopt $\alpha =
-1.11$, whereas our value is closer to $-1.6$. Similarly, Gronwall \&
Koo (1995) derived faint-end slope of the luminosity function that was
somewhat steeper than that of Metcalfe et al. (1991) for the bluest
galaxies, but otherwise quite comparable to that of Marzke \etal
(1994). The luminosity function for the latest galaxy types (which
dominate the cumulative luminosity function at low luminosity) from
Marzke \etal (1994) is described by $\alpha = -1.87 \pm 0.2$.

Why then do counts predicted from recent {\it observational}
determinations of the $K$ band luminosity function also roll over
around $K=22$ for q$_0=0.5$ (Mobasher \etal 1993, Glazebrook \etal
1995, Gardner \etal 1997)? None of these surveys contain low
luminosity galaxies in sufficient number to constrain the faint-end
slope. For example, Mobasher \etal use no data fainter than
$M_K\sim-23$ (only $\sim 0,1$ L$^*$) to derive their luminosity
function, while Glazebrook \etal simply {\it adopt} $\alpha = -1$. It
is relevant to note that these two surveys derive different $M^*_K$
and $\phi^*$ (by roughly a factor of 2 for each parameter, but in the
opposite sense) such that the bright end of the predicted counts match
the observations. Counts at bright magnitudes are little affected by
the value of $\alpha$. For example, the effective $M^*_K$ and $\phi^*$
of our local sample also fit the bright end of the counts ($K<17$), as
illustrated in Figure 6, even though our value of $\alpha$ is
considerably more negative. In contrast, $M^*$ and $\phi^*$ little
affect the {\it slope} of the faint end of the counts, and only
$\phi^*$ affects the count normalization.

In general, for steep luminosity functions, the slope of the counts
becomes increasingly dependent on $\alpha$ at fainter magnitudes. As a
consequence of the steep faint-end slope of the luminosity function of
our local sample (Bershady \etal 1998), the {\it model} counts at
faint magnitudes are increasingly dominated by low-L galaxies at
relatively low redshifts. This in turn makes the {\it slope} of the
faint end of the $K$ band counts insensitive to q$_0$ as well as the
cosmological constant. Yet Figures 4 and 6 do show that the count
amplitudes differ significantly (by a factor of $\sim$2 at $K=23$)
between q$_0$ = 0 and 0.5. {\it For a steep luminosity function slope,
it is the amplitude of the counts that is most affected by
cosmological parameters.}

Djorgovski \etal (1995) comment, however, on their inability to use
even the amplitude of the faint $K$ counts to constrain geometry
because of its sensitivity to many model parameters, including
evolution. This sensitivity is well illustrated in the simple models
presented by Koo (1990). Our analysis with the 1/V$_{max}$ models and
those of Gronwall \& Koo (1995) in the previous section indicates that
the {\it slope} of the faint end of the luminosity function is
critical in determining not only the count slope but the count
amplitude as well. Hence, even ignoring the effects of evolution,
there is no aspect of the faint counts (slope or normalization, either
in the optical or near-infrared infrared) that can be used reliably as
a cosmological probe without firm knowledge of the faint end of the
galaxy luminosity function and its evolution.

\section{Size, Color and Evolution of Field Galaxies}

\subsection{The excess revealed: counts as a function of image size}

\placefigure{fig7}

We can place additional constraints on galaxy evolution and cosmology
by exploiting our size information. In particular, galaxy size can be
used to estimate luminosity in special circumstances. Recall that
there is less than a 20\% change in apparent size for $0.8<z<3.5$ and
0.1 $<$ q$_0$ $<$ 0.5. To $K=23$, our 1/V$_{max}$ model predicts that
$\sim$75\% of the galaxies will be at such redshifts or higher, and
this fraction is only weakly dependent on q$_0$. Locally, galaxy
luminosity is observed to be tightly correlated with size. If this is
not a surface-brightness selection effect in local samples, and {\it
in the absence of size or luminosity evolution}, then apparent size
should continue at large redshifts to correlate strongly with
luminosity. Properly calibrated, size could be used to estimate
luminosity in this regime (modulo cosmological assumptions). Hence the
relative excess and slopes of small ($s$) and large ($l$) galaxies may
yield clues to the nature of the excess for the entire sample.

We start this analysis by taking advantage of the 1/V$_{max}$ models,
which transform various information, including sizes, from the input
sample into the output simulations. Full account of the PSFs are
included. More specifically, each object's size and shape (i.e. image
concentration) is compared to a grid of models appropriately smoothed
to derive the transformation. These transformations are expected to be
accurate and free of substantial bias because 1) the input radii are
large enough that the transformations depend weakly on object shape
and vary slowly with object size; 2) the input sample, taken in
0.8-1.2 arcsec seeing (FWHM) and at a median redshift of 0.13, is
typically better resolved than the output sample; 3) we have an
accurate estimate of both input and output PSFs.

Figure 7 compares the expected size distribution from the model (for
q$_0$=0 only) to that of the data, in the form of differential counts
for small ($s$) and large ($l$) objects. The shaded areas indicate the
variations in the observed counts if the size delimiter between $s$
and $l$ is varied by $\pm$10\% (spanning 30\% of the dynamic range in
size), and show that the qualitative behavior of the counts is not
sensitive to such variations. Our 1/V$_{max}$ model predictions of the
$M_K$ and $z$ distributions to $K=23$ are tabulated in Table 3 for
q$_0$=0. In the absence of strong evolution, large ($l$) galaxies
should correspond to L* galaxies observed at $z \sim 2$, while small
($s$) galaxies correspond to sub-L* galaxies typically at $z \sim 1$.
According to our findings in \S5, if these galaxies are mostly
disk-dominated systems, large galaxies should have half-light radii
(r$_{1/2}$) in excess of 3.5 h$_{50}^{-1}$ kpc, while small galaxies
will have r$_{1/2} < 3.5$ h$_{50}^{-1}$ kpc.

The observed slope for the $s$ counts is in good agreement with the
model until $K=22$ and $J=22.5$. Recall that we expect the estimate of
the $s$ counts to be too low because of the limited local sample for
the 1/V$_{max}$ model input (i.e. the 1/V$_{max}$ model
under-represents the true number of low luminosity objects even in the
absence of evolution). The relative increase in the excess for $s$
counts between $J$ and $K$ can plausibly be explained by
color-luminosity effects and redshift effects, both of which work in
the same direction to make the apparent colors of lower luminosity
objects bluer. The observed crossover point at $K \sim 22$ and $J \sim
23$ illustrates again why the counts do not roll over. According to
the 1/V$_{max}$ simulations, low-luminosity objects dominate the
counts at the faintest magnitudes, and are observed at relatively low
redshift where the differential volume is still increasing rapidly
with luminosity distance.

The surprising result is that the count {\it excess} is greatest for
larger ($l$) galaxies. This relative excess increases slightly in
$J$. The observed cross-over (where $s$ and $l$ surface densities are
equal) is about 1.7 mag fainter than the model predictions, while the
color at the cross over is as predicted ($J-K \sim 1$). Qualitatively,
the model difference in the slopes of the $s$ and $l$ counts matches
that seen in the data; however, the model amplitude for $l$ counts is
very low while the observed slope is relatively steeper.

\subsubsection{Comparison to optical observations}

We can check our results against two other recent studies that have
probed the size distribution of galaxies to comparable depth at
optical wavelengths. Smail \etal (1995) find a median half-light
radius that has an asymptotic value of 0.2 arcsec at $R \sim 26$,
based on ground-based images from Keck in comparable seeing to our
near-infrared data. This limit is comparable to $K = 23-23.5$, about
0.75-1.25 mag fainter than our cross-over point where our median
intrinsic $\theta_{0.5} = 0.44$ arcsec. For the most favorable case
(Gaussian profile), $\theta_{0.5}$ would correspond to a half-light
radius of 0.33 arcsec. At $R \sim 25$ Smail \etals Figure 4 indicates
their half-light radius is approaching 0.3 arcsec, and hence their and
our results are in close agreement.

An independent comparison can be made from the results of Roche \etal
(1996), who have measured sizes in the $I$ band from deep WFPC2
images.  Roche \etal measure a median half-light radius of 0.18-0.2
arcsec between $25 < I < 26$. This is comparable to Smail \etals
depth, if we adopt their median $R-I$ of 0.35 at $R=26$. Between
$23.5<I<24.5$, which corresponds closely to $K=22$ (cf. Moustakas
\etal 1997, Figures 8 and 9), Roche \etal measure a median half-light
radius of 0.25-0.32 arcsec. Hence their results agree with both the
results of Smail \etal and our own. Together, these three studies
indicate the size distribution of galaxies at these depths are
comparable when measured at wavelengths between 0.65 and 2.2
$\mu$m. As a consequence, the excess of apparently large, faint
galaxies observed in our deep near-infrared images should be found in
deep optical images as well.

\subsection{Constraints from near-infrared colors}

\placefigure{fig8a}

\placefigure{fig8b}

In addition to the amplitude and slope of the counts with and without
size information, colors can place constraints on the nature of the
excess galaxy population in these deep near-infrared images. The
usefulness of the single $J-K$ color available for our sample as a
redshift indicator is limited for field galaxies of diverse intrinsic
colors, although it has been attempted (Ellis \& Allen, 1983).  Figure
5 shows that synthetic $J-K$ colors for a range of observed and model
spectral energy distributions (SED) span a considerable range in
observed color at a given redshift. In particular, colors as blue as
$J-K$$\sim$1.5 are consistent with any redshift above 1, based solely
on observed SEDs. The redshift discrimination is worse if galaxy light
is dominated by stellar populations of very young ages. However, $J-K$
does offer significant leverage for discriminating redshifts of
galaxies with intrinsically red colors.

Figure 8(a) shows the $J-K$, $K$ color-magnitude diagram for galaxies
in our sample within the deepest portion of either the $J$ or $K$
images (i.e. within 0.5 mag of full depth). While it excludes part of
the sample used for counting, this sub-sample should be
representative. Figure 8(a) also extends to fainter magnitudes than
the counts. The total area sampled is $\sim 0.3 \times 10^{-3}$
degrees$^2$, and the total number of objects is 241. The median colors
to $K = 22.5$ for the $s$ and $l$ samples are listed separately and
combined in Table 4. The medians include objects with upper limits,
but are always bluer than the bluest of these upper limits. Note the
presence of a large number of galaxies fainter than $K$=21.5 with
colors near $J-K=1$ and the absence of many objects redder than
$J-K=2$. Qualitatively, this behavior is similar to the $I-K$ vs $K$
color-magnitude diagrams of Hogg \etal (1997) for $20<K<22.5$. We have
used their data to quantitatively check that, for all galaxies
combined, their trends in median $I-K$ with $K$ are similar to our
trends in median $J-K$ with $K$.

For comparison, Figure 8(b) illustrates one Monte Carlo simulation of
the expected color-magnitude distribution based on our non-evolving
1/V$_{max}$ models for q$_0$=0. This simulation matches the effective
observed area as a function of depth. However, the simulation does not
include the observed detection completeness, but instead is strictly
limited to $K<24$. (Note that the median model values listed in Table
3 use much larger simulations.) In both Figures, redshift tracks for
several fiducial galaxy spectra and luminosities are plotted; these
are discussed below.

The observed median $J-K$ color for small ($s$) galaxies is 1.6 mag,
with {\it no trend} in magnitude. While the lack of a trend is
consistent with the model prediction, the observed median color is
about 0.3 mag bluer than the model prediction (Table 3). If more lower
luminosity galaxies were included in the model input sample, the
resulting redshifts would become lower and the colors bluer. Hence it
is quite plausible that the $s$ galaxy population is consistent with
little to no evolution in an open universe.

On the other hand, the median $J-K$ color for large ($l$) galaxies is
1.6 mag averaged to $K$=22.5, but gets progressively bluer at a rate
of 0.1 mag per mag. By $K$=22.5, the observed median $l$ galaxy color
is over 1 mag {\it bluer} than the model predictions, and has the
opposite trend of color with magnitude. The observed median $l$ galaxy
color is also slightly bluer ($\sim 0.15$ mag) than the median $s$
galaxy color fainter than $K=20.5$, also in disagreement with the
models. In the models, the reddening trend for the $l$ galaxies is due
to luminous, {\it un-evolved}, early type galaxies seen at
progressively larger redshift with increasing depth (refer also to
Figure 5). The blueing trend is in no way anticipated by the
non-evolving 1/V$_{max}$ models.

\subsection{Evolution}

\subsubsection{What are the excess large, near-infrared-blue galaxies?}

We offer three explanations for the difference between observations
and the 1/V$_{max}$ models, two of which are plausible, and a third
which, while compelling for other reasons, we can rule out. First
there is the possibility of evolution, either in luminosity, color, or
both. There is now ample evidence, e.g., from the CFRS survey (Lilly
\etal 1995), that between $z=1$ and $z=0$ there has been modest
luminosity evolution for blue galaxies, with possibly a steepening of
the blue galaxy luminosity function, while there is little change in
the luminosity function for the redder galaxies. The form of this
evolution is qualitatively in the correct sense to explain the
increased number of blue galaxies. For example, $J-K$ for blue,
star-forming galaxies at $z$=1 is $\sim$1.5. However, this solution
would require also an evolution in size (decreasing with time), as
might occur in a scenario where galaxies at higher redshifts are
observed in the process of merging.  In addition, there should be
almost no very red galaxies at high redshift: While an L* galaxy
should be detectable to z$\sim$4 at $K=23$, there are few galaxies
observed with $J-K>2$, while the median $l$ model galaxy color at
$K=22$ is 3.4. As we discuss further below, this indicates we are
probing a redshift regime where ellipticals are observed at very young
ages.

Another possibility is that our sample is revealing a high space
density of low surface brightness dwarfs, detected here at relatively
low redshift ($z<1$). Holmberg (1975) pointed out that lower
luminosity galaxies tend to have lower surface-brightness. This trend
is seen in our input sample for the 1/V$_{max}$ models. What is
required to fit the deep $K$ observations is a substantially larger
dynamic range in surface-brightness for a given luminosity than
currently observed, along the lines of what has been suggested by
Ferguson \& McGaugh (1995) and plausibly demonstrated by McGaugh,
Bothun, \& Schombert (1995).  This situation would violate one of the
conditions for apparent size and luminosity to be well
correlated. Likewise, if galaxies are detected at sufficiently low
redshift, apparent size and luminosity are expected not to correlate.

This second scenario can be presented as two extreme possibilities:
For low surface brightness dwarfs to dominate the $l$ counts, they
will need to be at (a) $z<0.8$ for L$<$0.01L$^*$ or (b) $z<0.07$ for
L$<$0.0001L$^*$. The former case (a) cannot be ruled out by our data,
and moreover it is difficult to distinguish between this scenario and
one where high redshift, luminous, young galaxies contribute to the
excess, large blue galaxy population. The primary reason for this is
because their observed {\it blue} colors at disparate redshifts are
quite comparable, both in the optical and the near-infrared. This is
illustrated in Figure 8(b) for $J-K$, where two spectra for N4449 are
shown assuming a factor of 100 (5 mag) difference in luminosity.  The
two redshift tracks in this Figure differ in $J-K$ color by $\sim 0.5$
mag at a given apparent $K$ magnitude; between a redshift of 0.5 and
2, $J-K$ changes by $\sim 0.2$ mag.  Recall also that if sufficiently
young galaxies are present at any redshift, $J-K$ offers almost no
leverage for estimating redshift (Figure 5). The near degeneracy in
color and redshift are still worse in the optical, as illustrated in
Figure 9.

\placefigure{fig9}

For case (b), galaxies must be very red in the optical to
produce a median color of $J-K$=1.6 at such low redshifts.  Such an
abundant population of faint, red galaxies has yet to be directly
confirmed at low-to-intermediate redshift, but they have long been
suggested to exist as the ``end state'' of post star-forming galaxies
(e.g. Searle, Sargent \& Bagnuolo 1973, Huchra 1977, Babul \& Rees
1992). If true, this would be the first evidence for their
existence. At a limiting redshift of 0.07, there is only about 2
h$_{50}^{-3}$ Mpc$^3$ down to $K = 22.5$ accessible in our
survey. Hence if such objects dominate the counts at this depth for
large galaxies, their space density is of order 30 h$_{50}^{-3}$
Mpc$^{-3}$.

We can test this low redshift, red dwarf galaxy scenario (case [b]) by
checking the optical colors of our sample, taking advantage of the
$KG3$ and $I$ band data in one of our fields from Hall \& Mackay
(1984). While their catalogues are not as deep as ours, we have
matched 14 objects, only 2 of which are not in the deep portion of the
near-infrared images (both happen to be very red in $J-K$). Of the
remaining 12, plotted in Figure 9, six are brighter than $K=19$. Of
these, 1 is the central star with characteristically red $R-I$ and
blue $J-K$. Four of the others are consistent with intermediate galaxy
spectral types between $1<z<1.5$, ($0.9<R-I<1.2$) while the fifth is
very blue ($R-I=0.2$) and consistent with $2.5<z<3.5$ for a blue
galaxy spectral type. The remaining 6 are between $20<K<21.5$, and are
of the most interest since they are faint enough to sample the region
where there is a significant excess of large, blue galaxies. Two of
these are $s$ galaxies, with colors consistent with blue galaxy
spectral types between $1<z<3$. Of the four $l$ galaxies, three are
consistent with blue-to-intermediate galaxy spectral types between
$1<z<3$. A fourth is very red in $R-I$ given its $J-K$ color, and
falls between the elliptical track and the stellar locus. A sixth is
consistent with blue spectral type around $z=0.25$. In summary, there
is {\it no evidence} for any galaxies at $z<0.25$ to $K=21.5$, and
only one galaxy plausibly at $z<1$ with red optical colors.  The
majority (7 of 11) of galaxies have colors consistent with blue to
intermediate spectral types at $1<z<1.5$.  Moustakas \etal also find
that most of their faint galaxies have blue $V-I$ colors.
Hence, we can rule out the possibility that there exists a high space
density of {\it red} dwarf galaxies at low redshift {\it in the
field}.

\subsubsection{Optical confirmation of an excess of large, optically-blue galaxies}

Roche \etal (1996) also find an excess of galaxies with large
half-light radii ($>$ 0.4 arcsec) with blue optical colors ($V-I<1.2$) in
the range $22<I<24$. This excess is claimed with respect to their own no
evolution and pure luminosity evolution models which use a steep
luminosity function slope comparable to ours ($\alpha = -1.65).$ The
excess large galaxies, they note, appear morphologically to be spirals
and irregulars, but not ellipticals.  It is reasonable to assume that
Roche {\it et al.} are indeed observing the optical counterpart of the
excess population of large, {\it near-infrared} blue galaxies in our
survey. Does this information offer new clues as to the nature of
these galaxies?

Their interpretation is that this excess of large galaxies is due to a
1 mag brightening of today's L$^*$ galaxies by $z = 1$ to 2. However,
they also find a large number of small objects in the same magnitude
range and with similar colors (the distribution peaks at $V-I\sim0.8$
for both). Yet they interpret the small galaxies as dwarfs at $z<0.5$,
present due to their steep luminosity function slope. To support this
distinction between small and large galaxies {\it with the same
optical colors}, they marshal evidence from spectroscopic surveys
(e.g. Lilly \etal 1995) that indicate 1 mag of luminosity evolution
for bluer galaxies by $z=1$, and Cowie \etal's (1995) ``chain''
galaxies at $22<I<23$, which are large, blue and between $1<z<1.6$. On
the other hand, recent spectroscopic results from Koo \etal (1996)
identify the bulk of the galaxy population to $I<24$ as sub-L$^*$ with
a median redshift of $\sim$0.8.  Neither the depths nor the
completeness of the faintest spectroscopic surveys is sufficient to
determine quantitatively the contributions from low and high redshift
to the large, blue galaxy population observed at $I \sim 26$ or $K
\sim 23$.

An alternative scenario to pure luminosity evolution is the bursting
dwarf hypothesis of Babul \& Ferguson (1996). Roche \etal (1996)
dismiss this model because they claim the predicted size distribution
is too small. However, close scrutiny of Babul \& Ferguson's Figure 18
shows that the predicted size distribution to $I \sim 25$ is not
unreasonable, with a peak in the half-light radius distribution near
0.25 arcsec. What is more problematic for Babul \& Ferguson's model is
the $I-K$ vs. $K$ color-magnitude diagram, which shows far too few
blue galaxies by $K=22$ compared to, e.g. Moustakas \etal's Figure 8.
Moreover, bursting dwarfs are not needed to keep the $K$ counts
rising, and indeed a steep non-evolving luminosity function can match
the same observations, as we have demonstrated. The most important
test of the bursting dwarf hypothesis will be to see if the $K$ counts
for small objects rise more steeply for $K>23.5$ than illustrated in
Figure 7 for $K<23.5$.

Finally, it is worth commenting on the results of the infall formation
models calculated by C\'{a}yon \etal (1996), which predict that
smaller half-light radii should accompany luminosity evolution. If one
accepts the luminosity evolution scenario favored by Roche \etal
(1996), one would expect that there would {\it not} be an excess of
large, blue galaxies at faint magnitudes. Roche \etal (1996) dismiss
the alternative possibility that there exists a substantial population
of low surface-brightness dwarfs in their sample. They claim selection
effects would keep such objects out of their sample. However, at a
given {\it apparent} size and surface brightness, there is no
preference for photometrically detecting a galaxy at low or high
redshift. Perhaps, then, there is further reason to consider the
possibility that relatively low redshift, low surface-brightness
dwarfs contribute substantially to the large, blue galaxy excess
population. The question remains, therefore, how can low
surface-brightness dwarfs at relatively low redshifts ($z\sim0.5$) be
distinguished from and high surface-brightness giants at high redshift
($z>1$) when both have similar, extremely blue colors yet are beyond
the limits of spectroscopy? This question cannot be answered here.

\subsubsection{A deficit of red galaxies?}

While we have focused our attention on interpreting the nature of the
surfeit blue galaxies, particularly those with large image sizes, it
is relevant to consider whether the 1/V$_{max}$ models predict too
many red galaxies. An observed absence of red galaxies might indicate
that at least one source of the blue excess population are early-type
galaxies observed at high redshift when they were more luminous and
bluer in color. Because we can use $J-K$ instead of an optical or
optical-infrared color, we are much less sensitive to recent but small
(in terms of mass) bursts of star formation superimposed on old,
underlying stellar populations. In this sense, $J-K$ puts stronger
limits on the presence or absence of galaxies whose light is dominated
by old stellar populations (cf. Zepf 1997).

To $z \sim 3$, the non-evolving 1/V$_{max}$ model $J-K$ colors are
comparable to passively evolving models of present-day ellipticals
with formation redshifts ($z_f$) greater than $\sim 10$ (see Figure 5
from McCarthy 1993). Luminosity evolution, however, which is {\it not}
included in the 1/V$_{max}$ models, {\it is} appreciable for passively
evolving models even in the near-infrared. Luminosity evolution in the
$K$ band amounts to $\sim$ 0.5 mag at $z = 1$ and $\sim$ 1 mag at $z =
2$ in an open universe for passively evolving models of present-day
ellipticals with $z_f>4$. This means that we can set strong upper
limits on the number of red objects expected for passive evolution,
since sources will only brighten but not become bluer with
redshift. Hence the predicted number of red sources within our survey
limit will only increase in the passively evolving scenario with
respect to no evolution models.

Operationally, we define a ``red envelope'' in the $J-K$ vs. $K$
color-magnitude diagram [Figures 8(a) and 8(b)] as that region
redwards or brighter than the redshift track for an unevolving
elliptical with absolute magnitude near L$^*_K$.  It is important to
keep in mind that because there is no direct redshift information, at
any given apparent magnitude and color there is an ambiguous trade-off
between luminosity and redshift for a given SED. The ambiguity
increases when other SEDs are included. For this reason, we restrict
the analysis to luminosities above L$^*_K$ corresponding to the
reddest SEDs, since galaxies with bluer SEDs are expected to be
preferentially at lower luminosities (in the absence of evolution).
However, in order to check the sensitivity of our threshold, we also
define a second, more inclusive envelope using the redshift track for
the same red SED with absolute magnitude near 0.1 L$^*_K$.

Note that in the absence of evolution, some very luminous (L$_K > 10$
L$^*_K$) galaxies are expected in deep samples. The simulated objects
in Figure 8(b) at the reddest colors for a given magnitude are {\it
not} intrinsically ``ultra''-red, but simply over-luminous.  At great
depths one might expect even more luminous objects to be found: While
such objects are rare, more volume is sampled than in the input sample
of our 1/V$_{max}$ models. For example, Hu \& Ridgeway (1994) found
two ``red'' objects at $I = 18.5$ ($I-K=6.5$) which they believe,
based on $BIJHK$ colors, are unevolved ellipticals at $z=2.4$ and 10
L$^*$. However, Graham \& Dey (1996) find one of these objects appears
to have broad emission consistent with H$\alpha$ at $z=1.44$. If true,
this is indicative of intensive star formation or nuclear activity,
and the red colors indicate dust and not an old
population. Nonetheless, the {\it upper} envelope used by Elston \etal
(1988) in $R-K$ vs. $K$ (corresponding to M$_V$=-23.3, H$_0$=50,
q$0$=0), for example, would have excluded two galaxies at moderate
redshifts with luminosities 0.4 mag brighter but colors of present day
ellipticals from Bershady (1995). In the spirit of placing an upper
limit on the number of galaxies with old stellar populations, no upper
limit in luminosity or color is imposed on our selection of ``red
envelope'' galaxies.

The numbers of objects observed and predicted from the 1/V$_{max}$
models to lie above the L$^*_K$ and 0.1 L$^*_K$ red envelopes are
presented in Table 5 in three intervals of $K$ magnitude. For q$_0=0$,
the number of observed $>$L$^*_K$ galaxies is very close to
predictions for small galaxies, whereas the observed number is too low
by about a factor of 3 for large galaxies. The number of observed
$>$0.1L$^*_K$ galaxies is low by about 30\% for both large and small
galaxies. However these deficits are only a 1-2$\sigma$ result, given
the small total number of objects. If the $J$ band upper limits are
all assumed to lie above the adopted envelopes, there are no deficits
in the last magnitude bin ($21.5<K<23$). Excluding these upper limits,
the expected number of large $>$L$^*_K$ galaxies is higher than
observed at all magnitudes, while for $>$0.1L$^*_K$ the deficit is
only in the two fainter bins.

For q$_0$=0.5, the expected number in the same region of color and
magnitude is roughly the same for large ($l$) galaxies (which are
predominantly intrinsically red in the models), but much lower for
small galaxies (which are predominantly intrinsically blue in the
models). The reason for this is because, to first order, the smaller
volume to a given redshift in a critical universe is offset by the
smaller luminosity distance. Hence, at a given apparent magnitude and
redshift, one sees fainter in the luminosity function. For bluer
galaxies, however, the $k$-corrections are more favorable, and one
sees galaxies of comparable rest-frame luminosity at higher redshifts
where there is less volume. Hence in a critical universe, bluer
galaxies are diminished in number relative to redder galaxies at a
given apparent magnitude.

A tentative result is that our observations are inconsistent at the
$\sim 1.5 \sigma$ level with the expected number of old, luminous
(L$>$L$^*$) galaxies in the range of $1<z<3$. This result is
insensitive to cosmological assumptions.

Given our low number statistics, it is more fruitful to ask if the
observed deficit of red, luminous galaxies could be responsible for
the surfeit of blue galaxies, large and small.  For example, the
redshift track of an evolving model galaxy with present day colors of
an elliptical but with $z_f=5$ is shown in Figures 8(a) and 8(b). This
illustrates one possible way a ``red envelope'' galaxy would evolve
and become bluer with apparent magnitude (redshift).  From Table 5 we
estimate that only as much as 10-30\% of the large, blue galaxy
excess, and 20-30\% of the small blue galaxy excess can be made up in
this way in the range $20<K<23$. The percentage declines towards
fainter magnitudes. The remaining excess must therefore come from some
as of yet indeterminate combination of galaxies evolving to
$\sim$L$^*$ and blue colors, and a steep luminosity function, as we
have previously discussed.

\subsubsection{Evolution away from the ``red envelope''}

What other evidence is there that early-type galaxies have
substantially evolved at $z>1$? From other deep imaging surveys, for
example, Djorgovski \etal show an absence of objects with red $r-K$
for $K>21.5$. At somewhat brighter limits, Cowie \etal (1994, 1995)
also show an absence of objects with red $I-K$ for $K>20.5$. Cowie
\etal (1994) note that the surface density of faint, red ($I-K>4$)
galaxies is less than what would be expected in the absence of
evolution. Their comparison, though, is to the expected surface
density for all galaxies integrated to 8 mag fainter than L$^*$ (Gardner
\etal 1993).

There is, however, abundant recent evidence that red galaxies in
clusters become more luminous in the past. Luminosity evolution has
been inferred to $z=1.2$ using the Tolman test (Dickinson 1995, Pahre
\etal 1996, Schade \etal 1996), and to $z=0.4$ using Fundamental Plane
relations (Van Dokkum \& Franx 1996, Bender \etal 1996). In these
analyses, red galaxies are assumed not to have evolved in size. These
studies conclude that changes in luminosity with redshift are
consistent with passive evolution in clusters. Lilly \etal (1995) find
little evidence for luminosity evolution in the field for red galaxies
in the range $0<z<1$. Yet Schade \etal's (1996) study finds evidence
for luminosity evolution in both field and cluster ellipticals.

Evidence for color evolution in red galaxies has been less
forthcoming. In clusters, however, Aragon-Salamanca \etal (1993) found
that by $z=0.9$ few cluster members were as red in $R-K$ as present
day ellipticals. Recent results from Stanford \etal (1998) also show
that elliptical galaxy cluster members do appear to get bluer over the
range $0<z<0.9$, consistent with expectations of passive evolution and
large $z_f$. However, this amounts to $J-K$ getting bluer by less than
0.2 mag for cluster E/S0 galaxies over this redshift range. A more
dramatic form of evolution has been claimed by Kauffmann \etal
(1996). Upon reanalysis of Lilly \etal's data, they find 2/3 of the
galaxies ``earlier'' than Sa are gone by $z=1$, assuming passive
luminosity evolution and $z_f=5$ (H$_0$=50, q$_0$=0.5). Notably, Lilly
\etal (1995) performed a similar analysis and inferred no evolution in
color if no evolution in luminosity were assumed (consistent with
their own results for the luminosity function of red galaxies).

There is no doubt that {\it some} galaxies do exist at substantial
redshifts that are red enough to be consistent with old, evolved
stellar populations. In the field, Koo \etal (1996) find several such
sources to $z=1$ in optically-selected samples. A large fraction of
Westerbork mJy sources are optically identified (to $V<21.5$) as
unevolved giant ellipticals up to $z \sim 1$ (Kron \etal 1985,
Windhorst \etal 1986). Recently, one such source has been identified
at $z=1.55$ (Dunlop \etal 1996). However, what is not known from these
studies is just what {\it fraction} of the expected sources are still
consistent with what we have defined as the ``red envelope.''
McCarthy's (1993) summary of the optical counterparts to 3CR and 1 Jy
sources indicates that, for $z>1.5$, $J-K$ colors broaden and become as
blue as $J-K = 1.25$. This color is bluer than the median colors in
our sample. Some radio sources, however, are as red as our red envelope
in $J-K$ at least to $z=2.5$.

In summary, there is ample evidence for evolution in luminosity and
optical-infrared colors in cluster ellipticals up to $z=1$, but the
evidence for field and radio samples is less secure or at least the
evolution is less homogeneous. By $z>1.5$, color evolution manifests
itself in the near-infrared ($J-K$) for radio samples.  However, some
``red envelope'' radio galaxies are still found at higher
redshifts. Our results here are broadly consistent with the results
from the radio surveys.  However, constraints on the formation epochs
of early-type galaxies is far from secure from those data. McCarthy
(1993) argues, for example, that the $r-K$ and $J-K$ colors of radio
galaxies are together inconsistent with a single-burst model for the
star-forming history of radio galaxies. Regardless of the actual
physical scenario, independent observational evidence indicates that
our observation of a deficit of faint ``red envelope'' galaxies
(corresponding to early-types galaxies at $z>1$) is plausible.

\section{Summary}

Our deep Keck near-infrared images have reached surface densities of
$\sim$ 300,000 mag$^{-1}$ deg$^{-2}$ at $J=23.5$ (corresponding to
$K=22.75$), which is equivalent to optical counts at depths of $B =
27$ and $I = 26.5$ (Metcalfe {\it et al.} 1995, Smail {\it et al.}
1995, Williams {\it et al.} 1996). Our $K$ band data go somewhat
deeper, and reach a surface density of $\sim$ 500,000 mag$^{-1}$
deg$^{-2}$ at $K=23$. This surface density is higher than any
published ground-based $B$ counts, and within 50\% of the surface
densities for the Hubble Deep Field in $V$ and $I$ bands (Williams
{\it et al.} 1996).

A robust result, not dependent on color or size, is that the $K$ band
counts do not roll over by $K=22.5$. The same is true for the $J$ band
at comparable surface densities. 

The excellent seeing conditions have allowed us to use size and color
together to identify the dominant galaxy type contributing to the
counts. By the faintest magnitudes, the smallest galaxies (i.e. within
the bottom 50\% of the size range for galaxies to these depths) begin
to dominate the counts, and have a median $J-K$ of $\sim$ 1.6. This
trend in size and $J-K$ color of the smallest galaxies is
qualitatively anticipated from our no-evolution models based on an
empirically-determined, local field galaxy luminosity function.  Such
galaxies, according to our models, correspond to relatively low
luminosity (L$<$0.1L$^*$) galaxies at $z < 1$; their abundance is due
to a relatively steep faint end slope for the $K$-band luminosity
function, dominated by blue galaxies at low luminosities.

As long as the volume density of galaxies continues to rise at lower
luminosities, the counts should continue to rise and be relatively
insensitive in {\it slope} to the cosmological volume. Even in the
absence of evolution (though strong evolution is likely at these
depths), improved measures of the faint end of the luminosity function
must be obtained before galaxy counts can be used as a cosmological
probe.

We have also been able to isolate the dominant galaxy type
contributing to the count excess, as reckoned with respect to mild or
no-evolution models. The most striking result of our deep
near-infrared survey to $K$=23 is that there is a substantial excess
of apparently {\it large} galaxies (i.e. within the top 50\% of the
size range for galaxies at these depths), compared to models with no
evolution. These galaxies are very blue in $J-K$ if they are at
intermediate or high redshifts ($z>0.5$), but relatively red if they
are at very low redshift ($z<0.25$). Hence this implies either (i) a
``new'' population of low redshift, low surface-brightness,
low-luminosity red galaxies, (ii) an intermediate redshift population
of moderately blue, diffuse galaxies with luminosities in the range
0.001-0.1 L$^*$, (iii) a strongly evolving population of galaxies at
high redshift observed at L$\geq$L$^*$, or (iv) some combination of
the three. We can rule out the first option based on the blue optical
($R-I$) colors of a random subset of our sample. However, weighing the
contributions from the second and third options is not possible
without spectroscopy or perhaps detailed morphological information
provided from higher resolution images.

We also find a relative paucity of very red galaxies compared to
models with no evolution. This deficit is in a region of $J-K$ color
and $K$ magnitude corresponding to early-type galaxies brighter than
L$^*$ at $z>1$ in the models. If it is assumed that such distant
galaxies have evolved to have bluer colors, then they can account for
no more than 30\% of the excess blue galaxies, large and small. The
result is insensitive to the assumed value of q$_0$. Further study of
the absence of these ``red envelope'' galaxies at faint magnitudes
should provide constraints on the epoch of early-type galaxy
formation.

\acknowledgments

We would like to thank W. Harrison for excellent support and
assistance with the NIRC camera, G. Wirth for his collaborative effort
in investigating new size statistics for faint object work,
C. Gronwall for access to her models. S. Majewski for access to CCD
data in Herc-1, and P. Hall \& C. McKay for access to CCD data and
catalogues in SA~57. MAB acknowledges support from NASA grant
NAG5-6032 and faculty research funds from Penn State University and
University of Wisconsin, Madison. MAB and JDL acknowledge support
from NASA through grant numbers HF-1028.02-92A and HF-1048.01-93A
respectively, from the Space Telescope Science Institute, which is
operated by the Association of Universities for Research in Astronomy,
Incorporated, under contract NAS5-26555. DCK acknowledges support from
NSF grant AST-88-58203 and faculty research funds from UC Santa Cruz.

\clearpage

\appendix
\section{Construction of Differential Counts in Images of Non-Uniform Depth}

The following is a step-by-step calculation of the counts in Table 1
for one case: SA~57, $K$ band, $s$-type. In general the entire
available area of each final image is used (i.e. Figures 1 and
2). These images are mosaics of many frames with a range of
offsets. Consequently the depth within the image is not uniform and
must be accounted for. This is necessarily a complicated calculation.

1. Raw counts (N): The raw counts are first tabulated in 0.5 mag
intervals, as listed in Table A1 (columns 1 and 2). The magnitudes
define the center of the intervals. The counts, when extracted from
the catalogue for each image, are truncated {\it not} at a fixed
magnitude, but at a fixed S/N such that the faintest counts come only
from the deepest (and smallest) area. That is, instead of: $$ m <
m_{50} $$ we specify $$m - 1.25 \ log(t/t_{max}) < m_{50} $$ where $t$
is the average value of the exposure map for pixels within the object
aperture, and $t_{max}$ is the maximum value in the image. Hence
$m_{50}$ is the magnitude limit (corresponding to 50\% detection) in
the deepest part of the image.

One further condition is made: In the deepest part of the image, we
count 0.5 mag fainter than in the other parts of the image. This is
accounted for when calculating the effective area (below in step
3). The point of this is to push the counts as deep as possible at
full depth, while not introducing noisier detections at brighter
magnitudes from the outskirts of the image.

2. Completeness function (d$f$): This is determined for the deepest
part of the image, as listed in Table A1 (column 3). The quantity
$\Delta$(d$f$) (column 4) is the estimated uncertainty in the
measurement of d$f$ based on scatter in simulations of detection
completeness. The index $j$ (column 5) is referred to in 3b) below.

3. Area ($S$): Here, areas are calculated for each depth. This accounts
for the fact that the sample is cut at fixed S/N.

a) We start with an image representing the square root of the exposure
map of the data image. The former has units of the square root of the
number of frames, $\sqrt{t}$, contributing to each pixel in the final
data image. The maximum value is $\sqrt{t_{max}}$ = 9.35, i.e. about 88
frames. Then we count the number of pixels down to $\sqrt{t} \sim
0.32$, i.e. 1/10th of a frame, in steps of $\sqrt{t}$ corresponding to
0.5 mag steps in limiting depth. (Less than 1 frame can contribute to
a pixel because the frames are registered on a sub-pixel level.) The
lower limit is somewhat arbitrary but unimportant because the area
converges. (For example, the number of pixels between $0.01 < \sqrt{t}
< 0.37$ is 515, or 0.5\% of the total area). The pixel counts ($N_{pix}$)
in these intervals are listed in Table A2 (columns 1-3, respectively).
The index $i$ (column 5) is referred to immediately below.  The
magnitude difference, d$m$, between the deepest and the $i^{th}$
interval is listed in column 4.

b) For each magnitude interval $m$, we sum up the areas in Table A2
starting with the deepest, and ending when $m$+d$m$ corresponds to the
last entry in Table A1 (the completeness file, step 2 above) or the
last entry in Table A2, whichever comes first. To account for how we
counted in step 1, only the faintest magnitude interval may use the
last entry in the completeness file. Hence the total area $S_{tot}$
is:
$$ S_{tot}(m(j)) = \sum_{i=1}^{min(i_{max},j_{max}-j)} N_{pix}(i),
\ \ \ \ j<j_{max},$$ and 
$$ S_{tot}(m(j)) = \sum_{i=1}^{min(i_{max},j_{max}-j+1)} N_{pix}(i),
\ \ \ \ j=j_{max},$$ where $j$ is the index in column 5 of Table A1,
$i$ is the index in column 5 of Table A2; $j_{max}$ and $i_{max}$ are
the maximum values (respectively 12 and 7). $S_{tot}$ is listed in
column 2 of Table A3 in units of square degrees assuming a plate scale of
0.1515 arcsec/pixel.

However, the effective area to be used for the counts is weighted by
the fractional detection at each depth: $$ S_{eff}(m(j)) =
\sum_{i=1}^{min(i_{max},j_{max}-j)} N_{pix}(i) \ {\rm d}f(j+i-1), \ \
\ \ j<j_{max},$$ and $$ S_{eff}(m(j)) =
\sum_{i=1}^{min(i_{max},j_{max}-j+1)} N_{pix}(i) \ {\rm d}f(j+i-1), \
\ \ \ j=j_{max},$$ yielding the results in Table A3, column 3. The
uncertainty in $S_{eff}$ arises from the uncertainty in d$f$. We
define ``boost'' to be $S_{tot}/S_{eff}$. The boost and its
uncertainty are in column 4 of Table A3.

4. The corrected counts (A) are calculated as $$ {\rm A} = {\rm N} \ /
\ ( S_{eff} \ {\rm d}m )$$ where d$m$ is the magnitude interval and
$S_{eff}$ is in square degrees such that A has units of counts
mag$^{-1}$ deg$^{-2}$. The corrected counts, A, in the original
magnitude bins are listed in Table A3, column 5. The uncertainties in
A are a quadratic combination of the statistical counting errors (1
$sigma$, Gehrels 1986) and the uncertainties in $S_{eff}$.  The
corrected counts averaged over two of the above 0.5 magnitude bins,
but stepped every 0.5 mag (1 original magnitude bin) are listed in
Table A4, column 2. Note again that A is corrected to counts
mag$^{-1}$ deg$^{-2}$. The log of these last values are what are
inserted into Table 1, as can be verified by inspection of column 3.

\clearpage

\begin{figure}
\plotfiddle{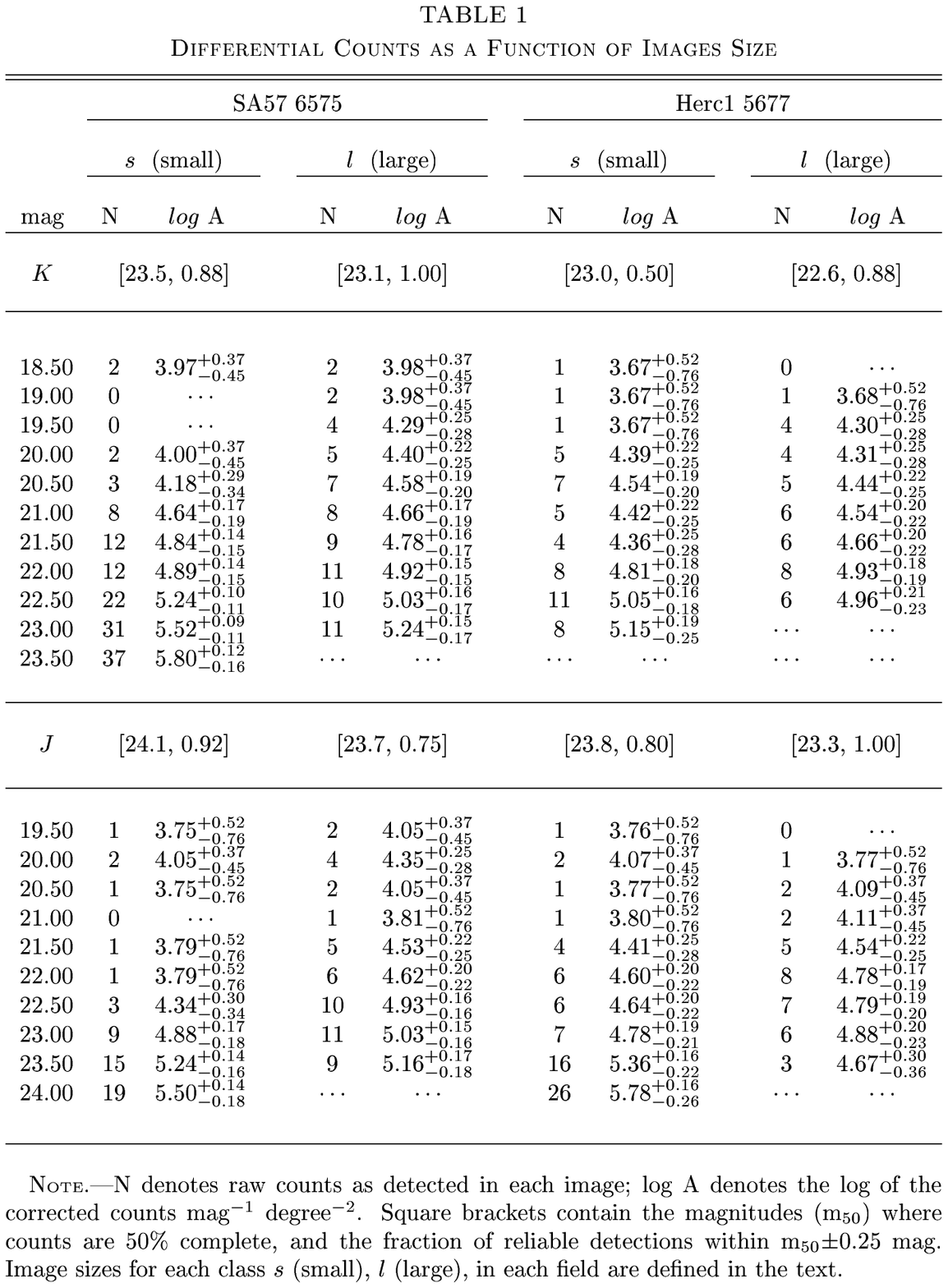}{8in}{0}{100}{100}{-300}{-100}
\end{figure}

\clearpage

\begin{figure}
\plotfiddle{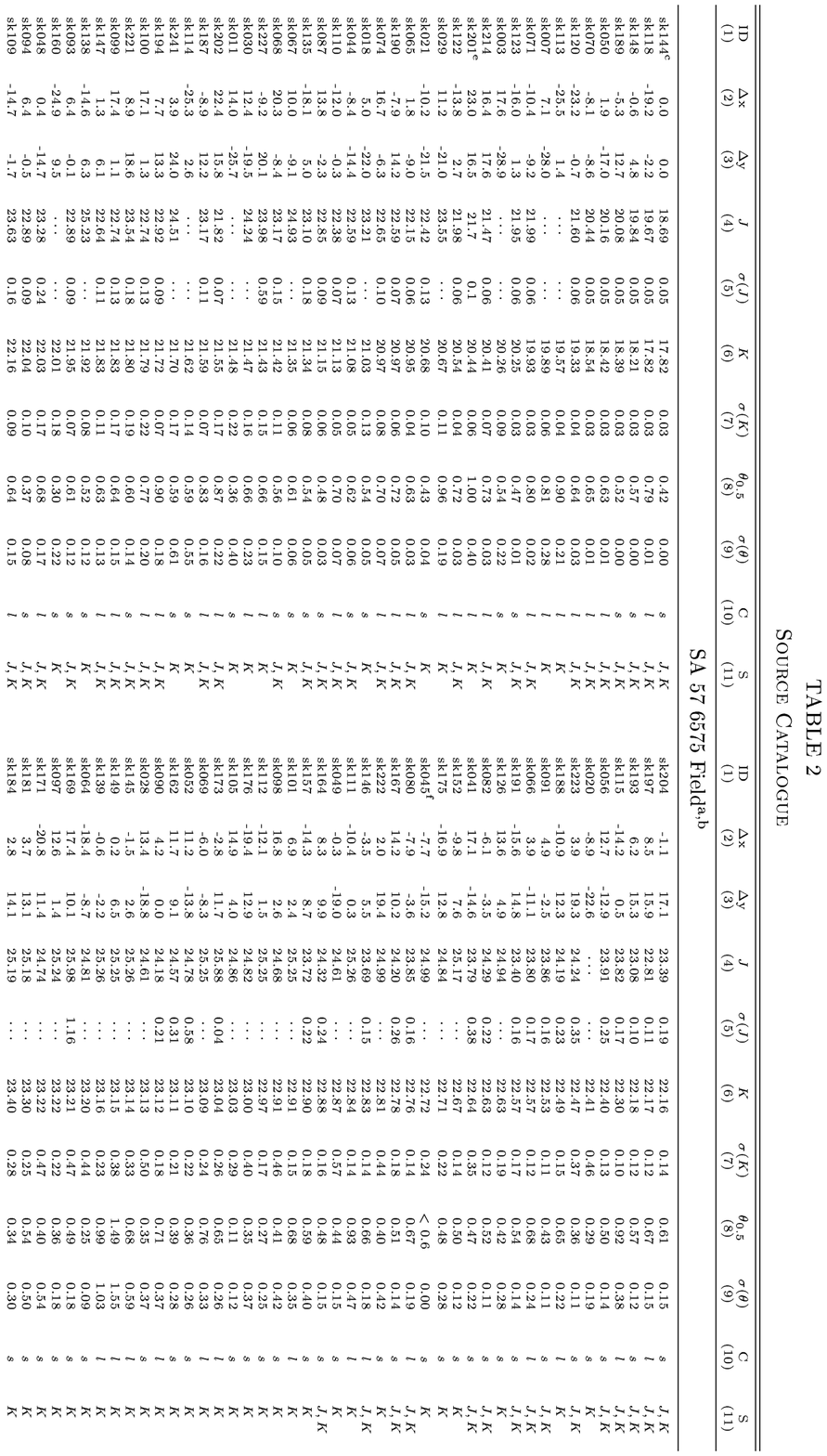}{8in}{180}{100}{100}{300}{675}
\end{figure}

\clearpage

\begin{figure}
\plotfiddle{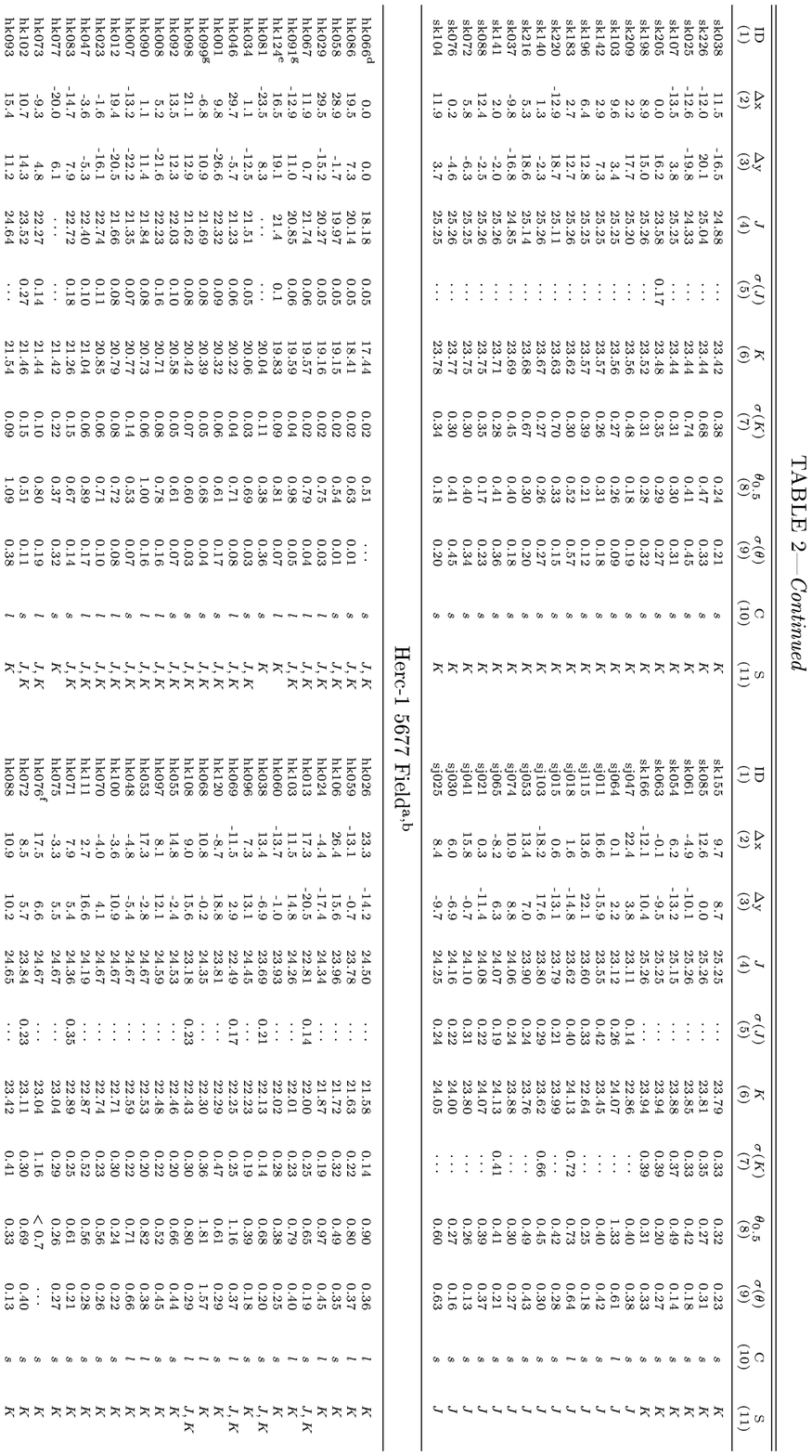}{8in}{180}{100}{100}{300}{670}
\end{figure}

\clearpage

\begin{figure}
\plotfiddle{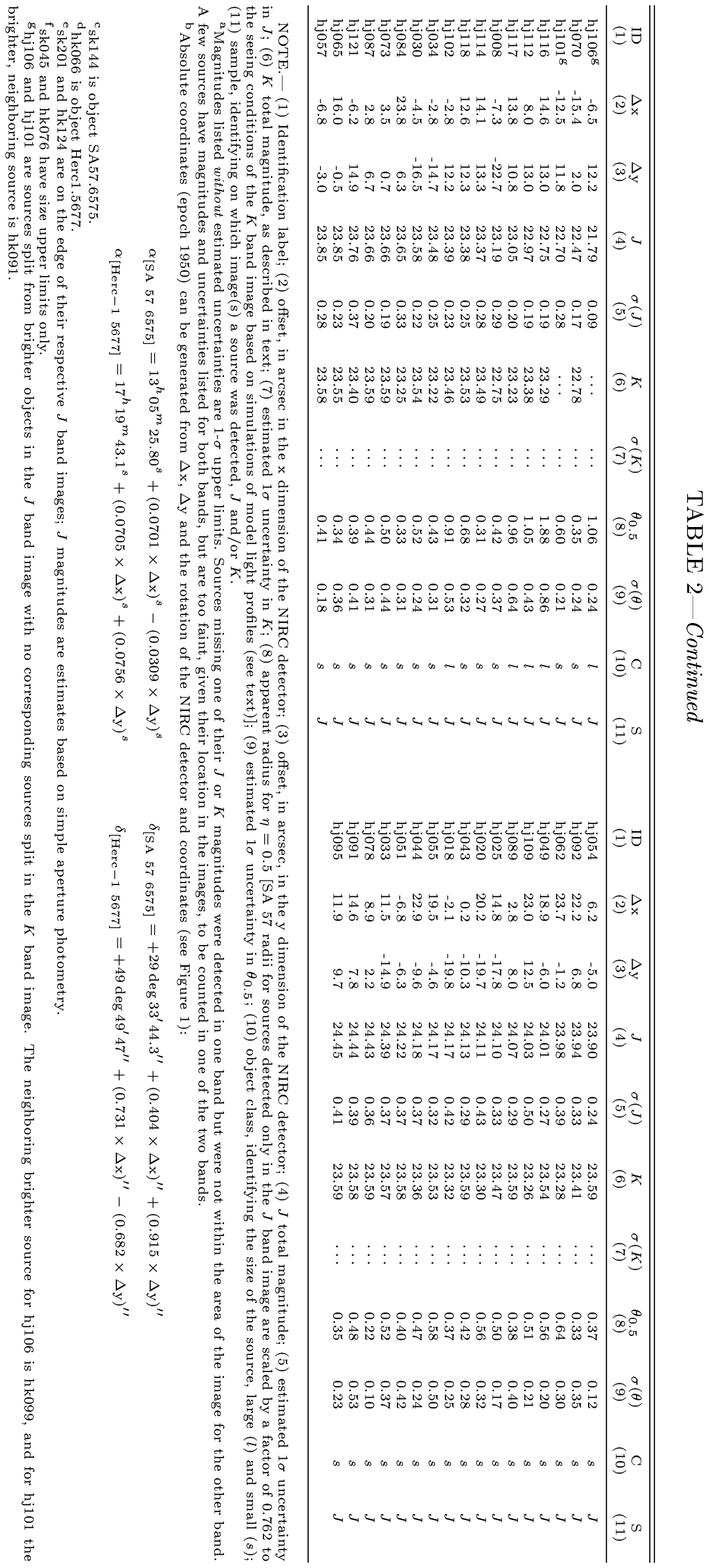}{8in}{180}{100}{100}{300}{670}
\end{figure}

\clearpage

\begin{figure}
\plotfiddle{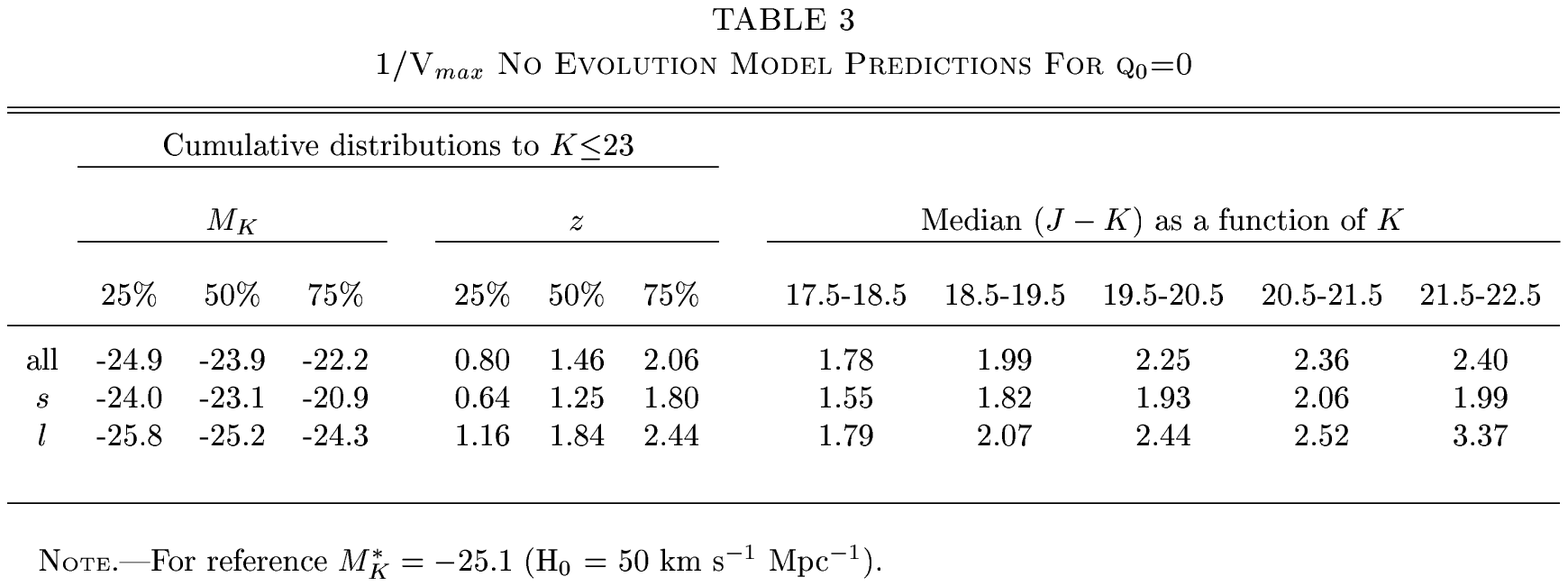}{4in}{0}{100}{100}{-300}{-250}
\plotfiddle{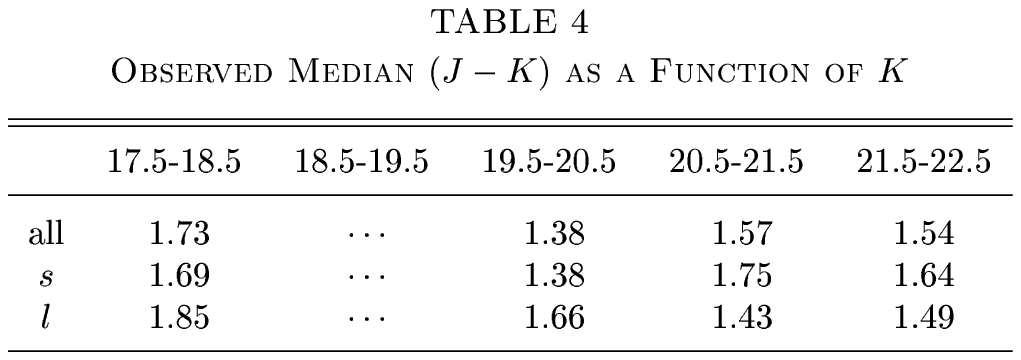}{4in}{0}{100}{100}{-300}{-200}
\end{figure}

\clearpage
\begin{figure}
\plotfiddle{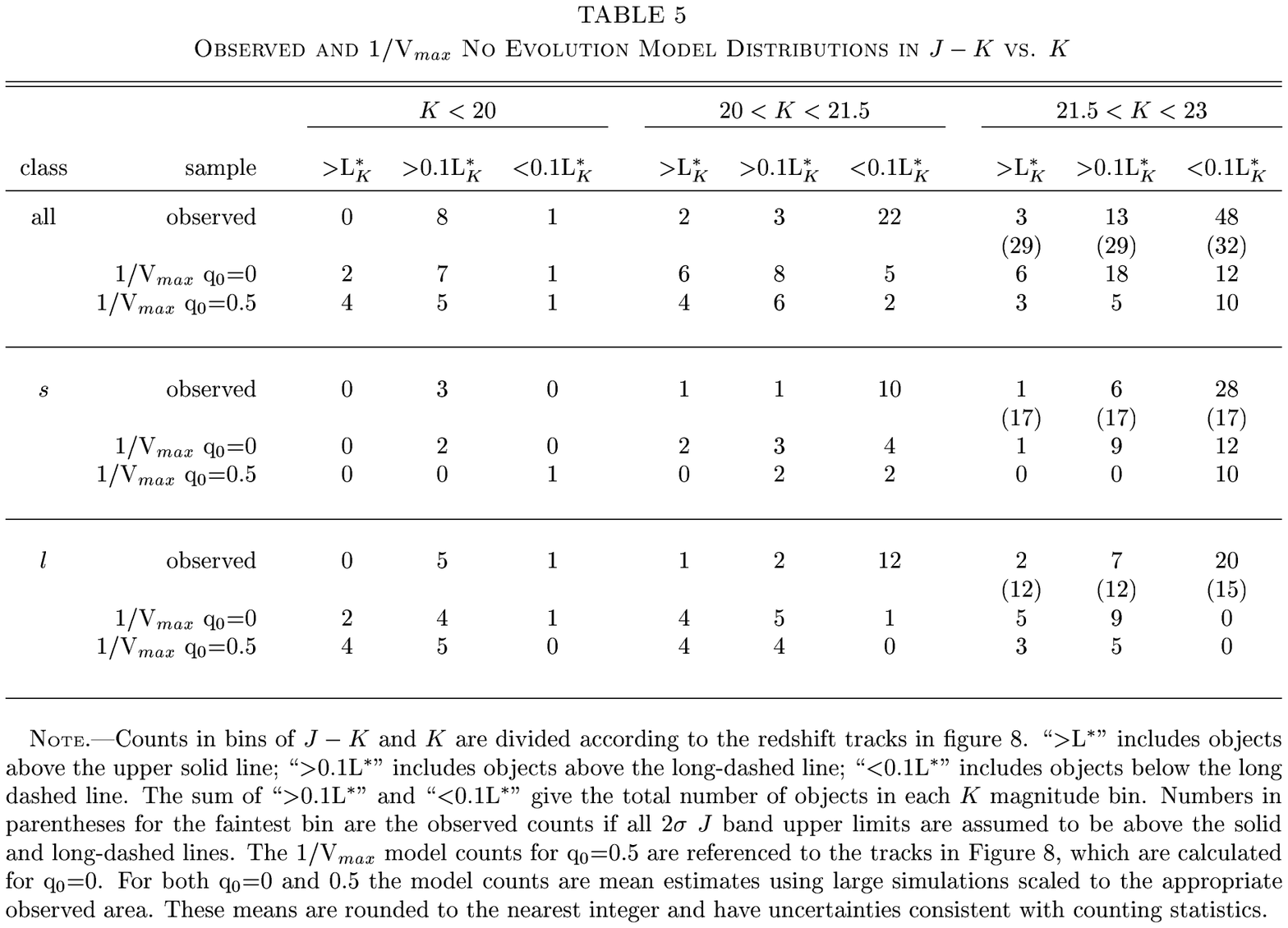}{8in}{0}{100}{100}{-310}{-50}
\end{figure}

\clearpage
\begin{figure}
\plotfiddle{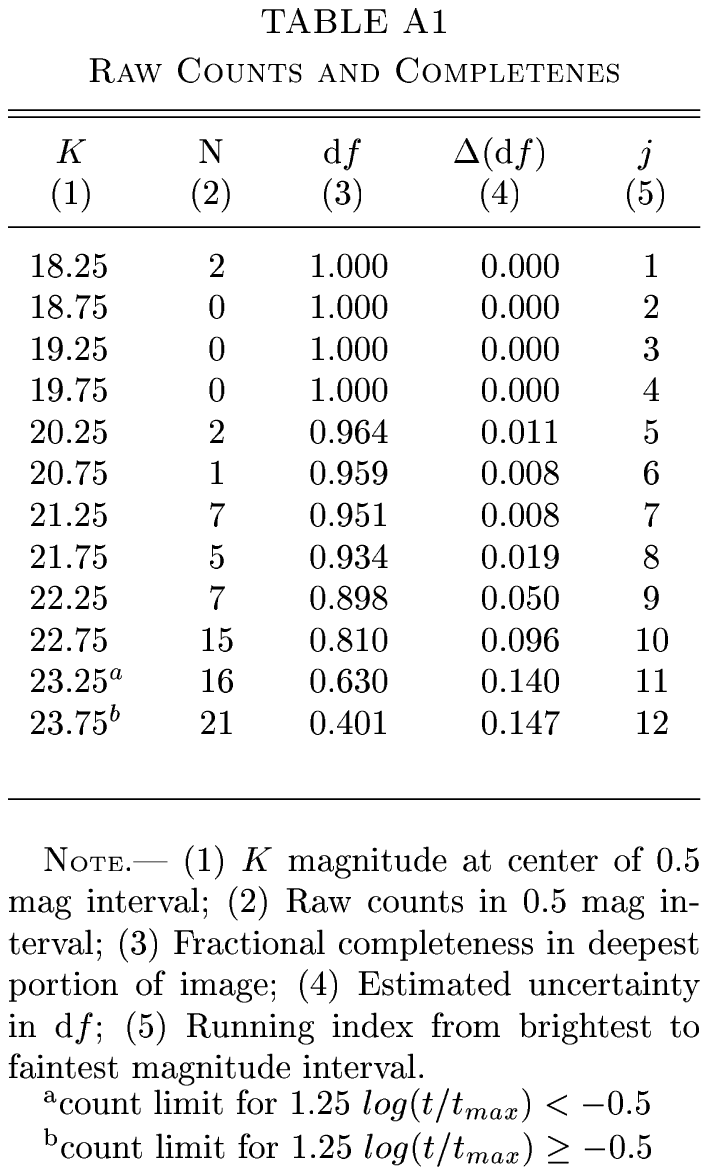}{4in}{0}{100}{100}{-300}{-250}
\end{figure}

\clearpage

\begin{figure}
\plotfiddle{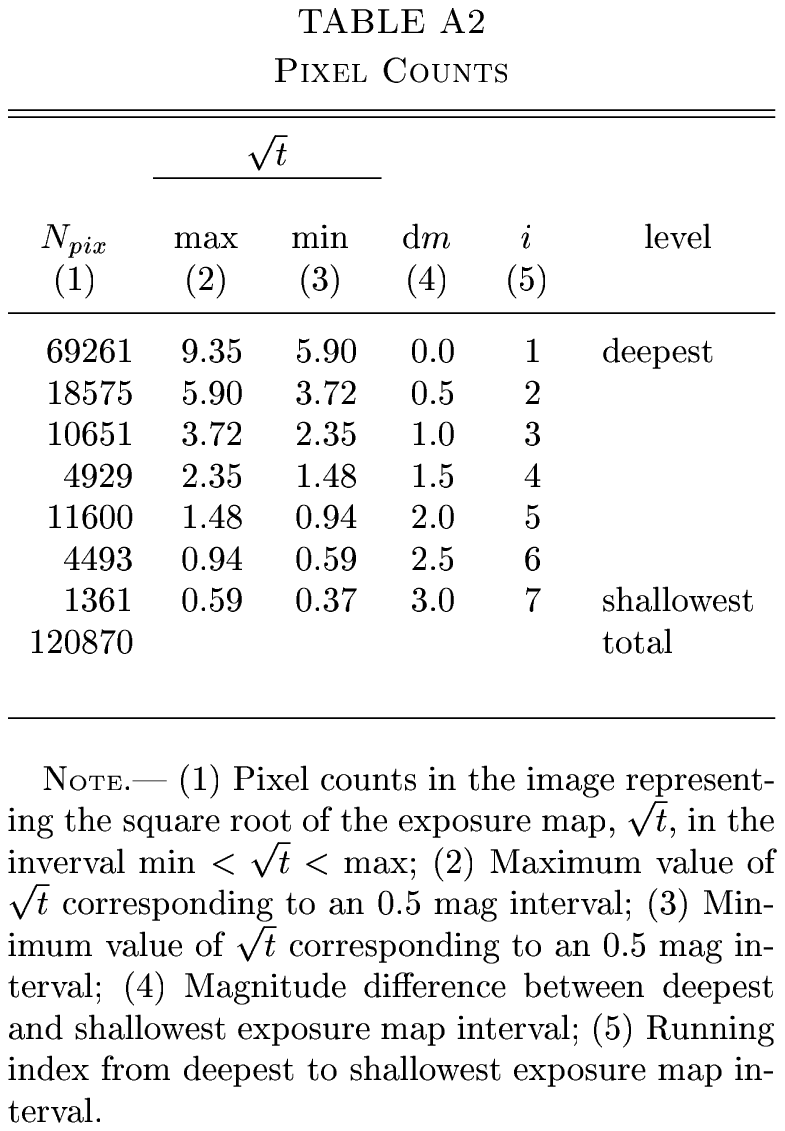}{4in}{0}{100}{100}{-300}{-200}
\end{figure}

\clearpage
\begin{figure}
\plotfiddle{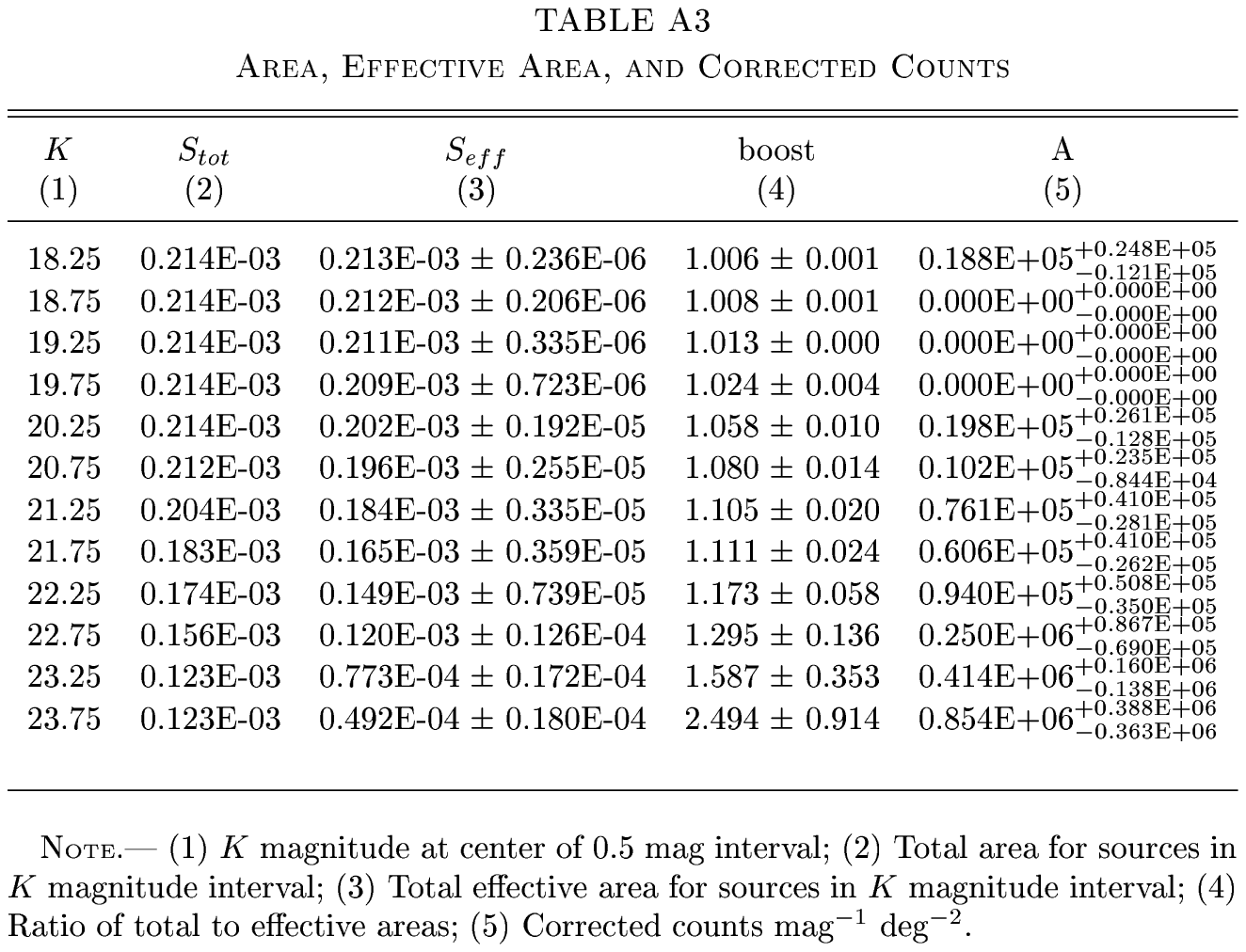}{4in}{0}{100}{100}{-300}{-225}
\plotfiddle{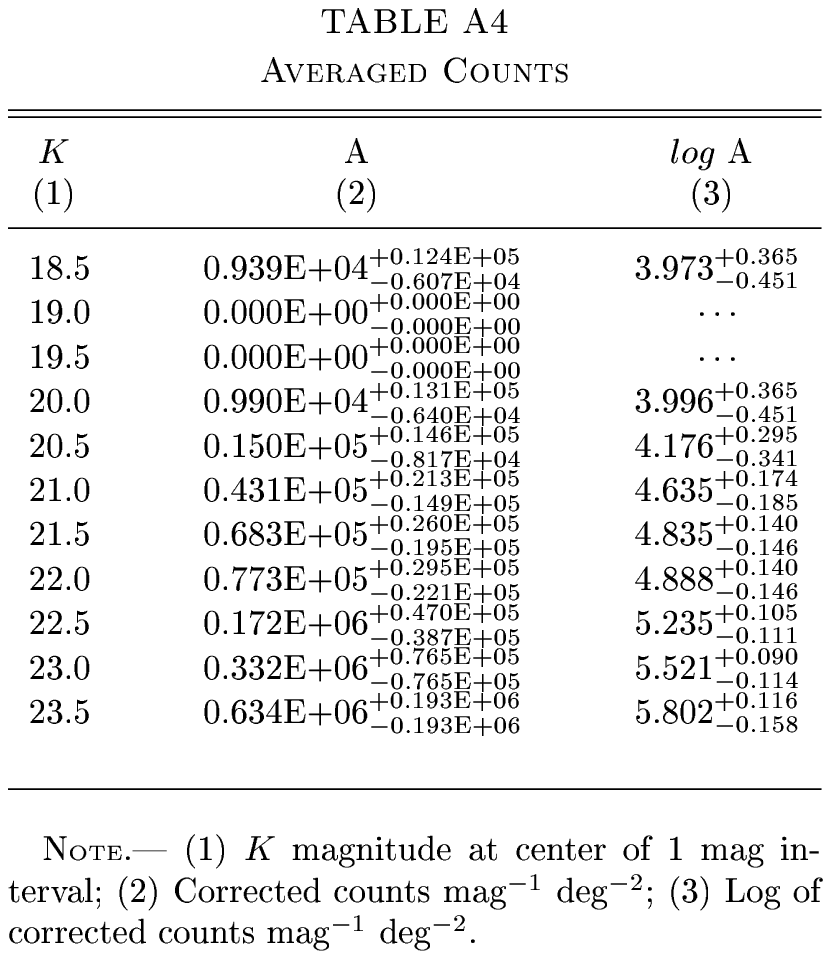}{4in}{0}{100}{100}{-300}{-250}
\end{figure}

\clearpage

\title {Figure Captions}

\figcaption[]{The $K$ and $J$ band images of the SA~57 6575 field
[plate 1] at the observed position angle of -23.8 degrees. Offsets (in
arcseconds) are referenced to the bright, central stellar source
SA57.6575 ($13^h05^m25.80^s \ +29\deg 33^{\prime} 44.3^{\prime\prime},$
1950), which has $V, I, J, K$ colors consistent with an M4V-M5V star
and $K$ = 17.82 $\pm$ 0.02 mag. Total exposure times were 20,880 sec
($K$) and 5,940 sec ($J$). Some fraction of data frames were rejected
because of unacceptable levels of detector noise. The remaining frames
were coadded to maximize the S/N for the reference point source as
described in section 3. The resulting useful exposure times are 17,500
sec ($K$) and 5,130 sec ($J$), and the coadded images yield FWHM of
0.54 arcsec ($K$) and 0.78 arcsec ($J$) for the reference stellar
source SA57.6575. The coadded frames displayed here have been scaled
by the square-root of their exposure map, which normalizes the noise
across the field (and consequently sources appear artificially faint
at the edges of the frame). These frames were used with FOCAS for
object detection, whereas photometry was done separately using the
FOCAS source list on the {\it original} coadded frames.\label{fig1}}

\figcaption[]{The $K$ and $J$ band images of the Herc-1 5677 field
[plate 2] at the observed position angle of -133 degrees. Offsets (in
arcseconds) are referenced to the bright, central stellar source
Herc1.5677 ($17^h19^m43.1^s +49\deg 49^{\prime} 47^{\prime\prime},$
1950), which has $V, I, J, K$ colors consistent with a K7V star and
$K$ = 17.42 $\pm$ 0.02. Total exposure times were 8,910 sec ($K$) and
2,970 sec ($J$). As for the images in figure 1, some fraction of data
frames were rejected, and the remaining frames were coadded to
maximize the S/N for the reference point source. The resulting useful
exposure times are 8,550 sec ($K$) and 2,880 sec ($J$), and the
coadded images yield FWHM of 0.64 arcsec ($K$) and 0.66 arcsec ($J$)
for the reference stellar source Herc1.5677. The coadded frames
displayed here have been scaled by the square-root of their exposure
map. Object detection and photometry were carried out as described in
the caption to figure 1 and in the text.\label{fig2}}

\figcaption[]{Fractional detection as a function of total $K$
magnitude and image size for the deepest image (SA~57 6575 field,
figure 1). Total magnitudes and sizes for ``small'' and ``large'' are
defined in the text. The lower-limiting case (``stellar'') has a 50\%
detection limit ($m_{50}$) roughly 0.4 mag fainter than for the
average ``small'' source.\label{fig3}}

\figcaption[]{Differential counts (number mag$^{-1}$ degree$^{-2}$) in
(a) the $K$ and (b) $J$ bands for all image sizes combined and
averaged over our two fields. Filled squares are corrected for
completeness and reliability; open squares are uncorrected. These are
compared to other $K$-band surveys (Moustakas \etal 1997 $[$M96$]$
Djorgovski \etal 1995 $[$D94$]$, Soifer \etal 1994 $[$S94$]$, Gardner
\etal 1993 $[$HDS/HMDS$]$, McLeod \etal 1994 $[$McL94$]$, and Sarracco
\etal 1997 $[$ESOK$]$); the models of Gronwall \& Koo (1995) for
q$_0$=0.05 (large-dashed line represents their best-fitting mild
evolution model including reddening, and the short-dashed line
represents their best fitting no-evolution model); and our 1/V$_{max}$
models for q$_0$=0 (solid line) and 0.5 (dotted line) with no
evolution (see text). The Sarracco \etal counts (ESOK) represent the
combined data for their two fields (ESOKS1 and ESOKS2).\label{fig4}}

\figcaption[]{Synthetic $J-K$ color vs. redshift for model and
observed SEDs (Bruzual \& Charlot 1993): (1) Unevolving colors for an
observed elliptical and NGC 4449 (heavy solid lines) and 16.4 Gyr
models for $\mu$=0.01 and 0.95 (light solid lines), where $\mu$ is the
fraction of galactic mass converted to stars each Gyr. H$_0$=50 km
s$^{-1}$ Mpc$^{-1}$ and q$_0$=0 is assumed throughout. (2) Evolving
colors for 16.4 Gyr models ($z_{formation}$$\sim$5) and for $\mu$=0.01
and 0.95 (dot-dashed lines). (3) Colors for constant age 0.01 Gyr
(dotted), 0.1 Gyr (short dash), and 1 Gyr age for a $\mu$=0.95 model
as they would be observed at each redshift. Note that only the
unevolving SEDs corresponding to today's ellipticals rise above
$J-K$=3. For reference, the color of a flat $f_{\nu}$ spectrum is $J-K
= 0.97$.\label{fig5}}

\figcaption[]{Differential counts (number mag$^{-1}$ degree$^{-2}$) in
the $K$ band for all published surveys, including this survey. M96,
D94, S94, HDS/HMDS, McL94, and ESOK surveys are defined in the caption
to figure 4. Other surveys include Mobasher \etal (1986, M86), Gardner
\etal (1993, HWS/G93), and Gardner \etal (1996, GSCF). Models are the
same as in figure 4.\label{fig6}}

\figcaption[]{Differential counts (number mag$^{-1}$ degree$^{-2}$) in
the $K$ and $J$ bands for small ($s$) and large ($l$) galaxies,
averaged over both fields. These are compared to our 1/V$_{max}$ model
simulations with no evolution for q$_0$=0. Image sizes and model are
defined in section 6.2. Shaded areas indicate changes to the counts
when the size delimiter between small ($s$) and large ($l$) is varied
by $\pm$10\% (dark shaded area for $s$ and light shaded area for
$l$).\label{fig7}}


\figcaption[]{(a) $J-K$ vs. $K$ for objects observed in the deepest
portion (to within 0.5 mag of full depth) of either the $J$ or $K$
images (both fields). Objects are marked as coded
in the key. Objects detected in only one band have their other
magnitude and color calculated with 2$\sigma$ upper limits, and are
included to depths (in their detected band) 0.5 mag fainter than the
50\% detection limit for stellar sources. The shaded area represents
the 50\% detection limits in $J$ and $K$ for the $s$ and $l$-type
objects in both fields. Error bars are marked for all objects detected
in both $J$ and $K$ and for all drop-outs brighter than
$K$=22.5. Model redshift tracks are labeled and described in figure
8(b).\label{fig8a}}

\setcounter{figure}{7}

\figcaption[]{(b) $J-K$ vs. $K$ Monte Carlo simulation for q$_0$=0
based on our 1/V$_{max}$ empirical models for the same area as the
observed sample in figure 8(a). Objects are marked as coded in the
key, with the determination of ``large'' and ``small'' based on the
same criterion as the observed sample (see text). Shaded areas
representing the 50\% detection limits of the observed sample are
repeated here. The simulation, however, is limited strictly to
$K<24$. Model redshift tracks are shown for an observed (non evolving)
elliptical galaxy SED near $M^*_K$ (top solid line, $M_K = -25$,
H$_0$=50, q$_0$=0); the same galaxy spectrum, 10 times brighter
(short-dashed line, $M_K = -27.5$), and 10 times fainter (long-dashed
line, $M_K = -22.5$); a blue, star forming galaxy near $M^*_K$ (bottom
solid line, N4449, $M_K = -25$); the same galaxy spectrum, 100 times
fainter (dotted line, $M_K = -20$); an evolving $\mu = 0.95$ model
with a present day age of 16.4 Gyr and $M_K=-25$ (dot-dashed line,
Bruzual \& Charlot 1993). Redshifts of 1, 2, 3, 4 and 5 are marked
with small squares and labeled for $z=1,3$ for $M_K\leq-22.5$ tracks;
redshifts of 0.25, 0.5, and 1 are marked and labeled for the
$M_K=-20$ track.\label{fig8b}}

\figcaption[]{$KG3-I$ vs. $J-K$ for 12 objects matched in the
deepest region of our SA~57 field to the catalogues of Hall \& Mackay
(1994). $KG3$ is similar to, but broader than $R$. Symbols are coded
as in figure 7 and 8, with larger symbols for $K<19$, and smaller for
$20<K<21.5$. The stellar loci for Main Sequence and giant stars are
shown as long-short-dashed and dot-dashed lines respectively.  Redshift
tracks for non-evolving galaxies (solid lines, top to bottom) are for
an observed elliptical, $\mu$=0.1 model, and N4449. Redshift tracks for
passively evolving galaxies (short-dashed lines, top to bottom) are
for $\mu$=0.95 and 0.1. Models are described in the caption to figure
4. Redshifts are indicated at 0, 0.5, 1, 2, and 3, and connected
between different SEDs by dotted lines.\label{fig9}}

\clearpage

\begin{figure}
\plotfiddle{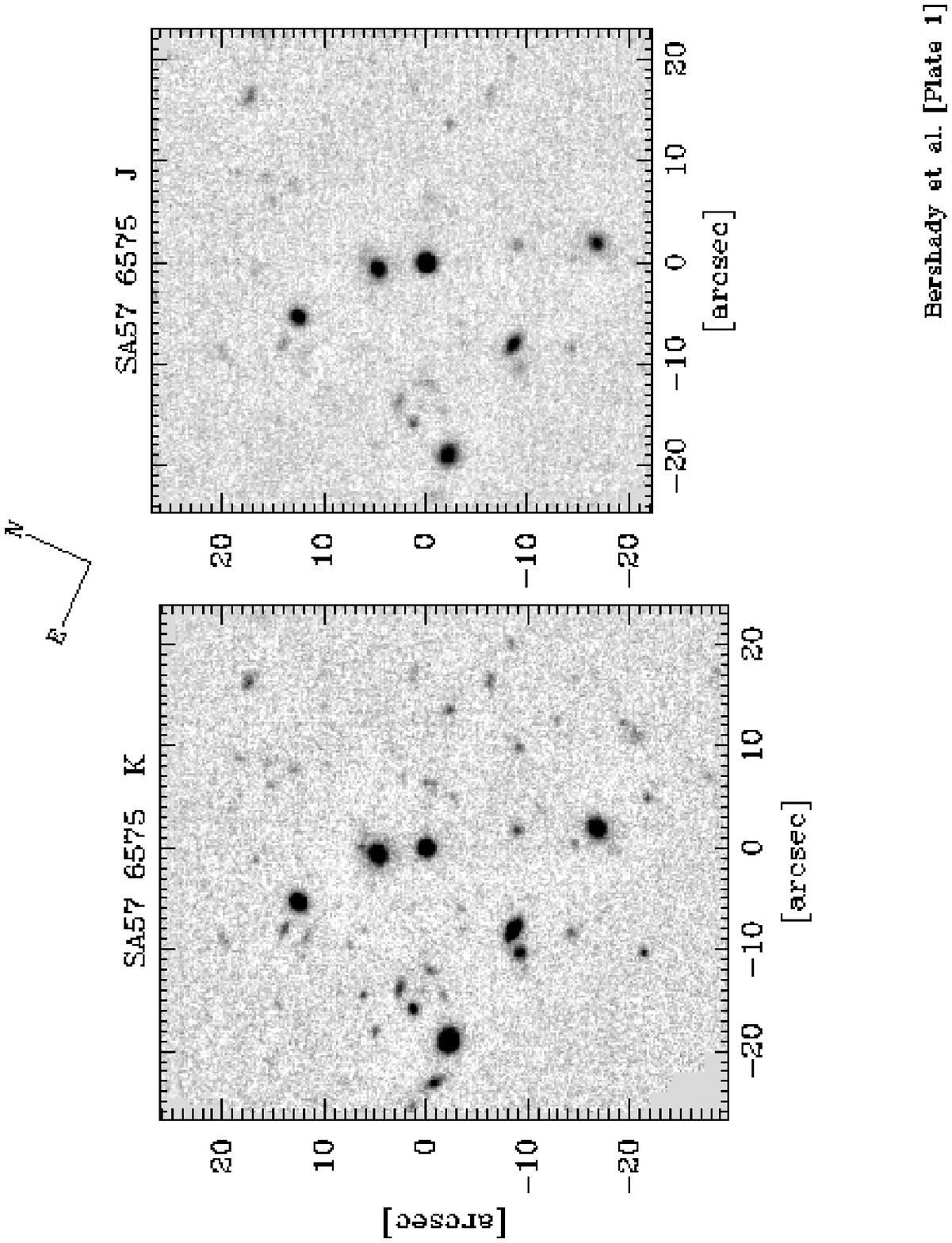}{8in}{0}{100}{100}{-300}{-100}
\end{figure}

\begin{figure}
\plotfiddle{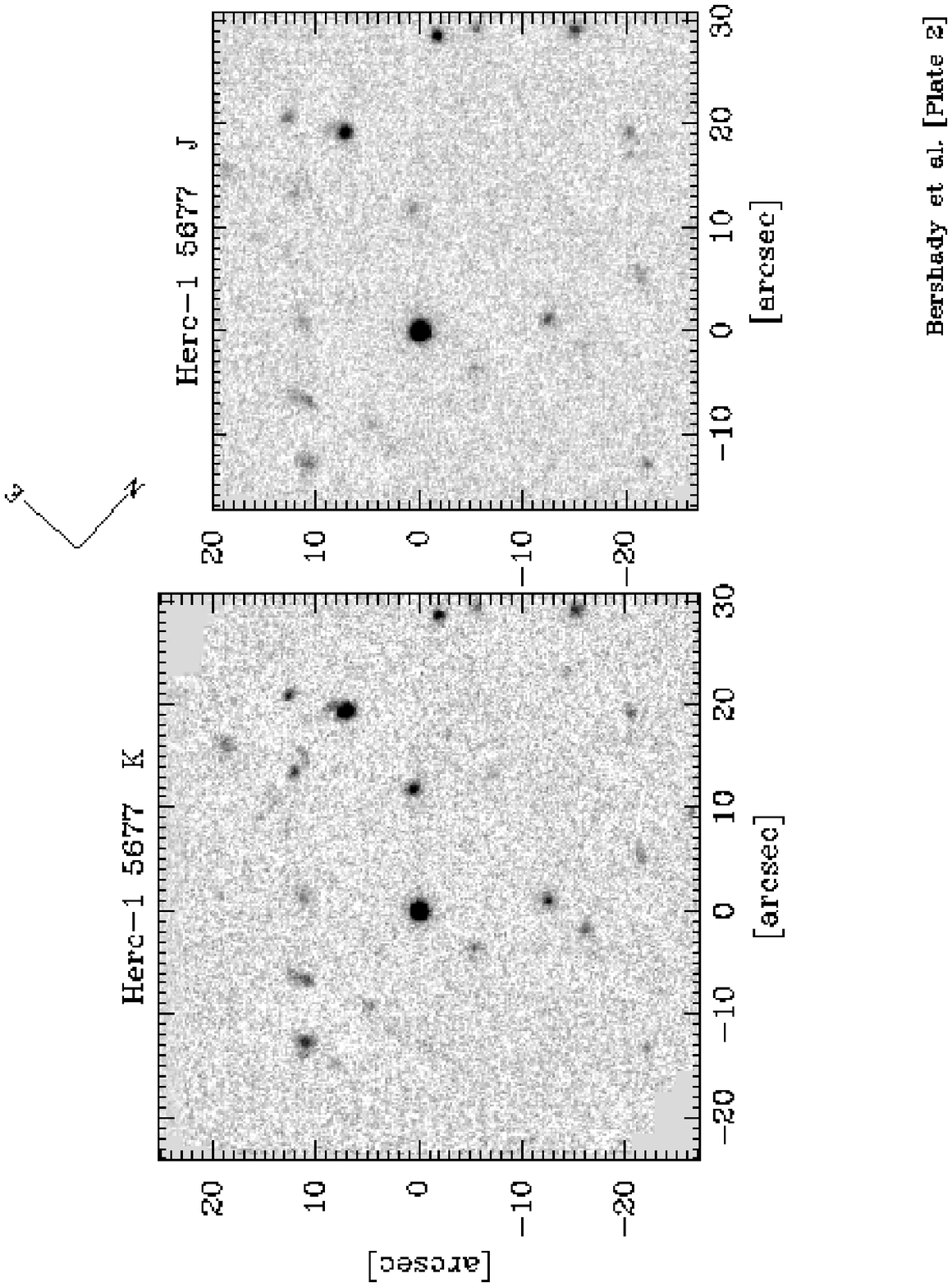}{8in}{0}{100}{100}{-300}{-100}
\end{figure}

\begin{figure}
\plotfiddle{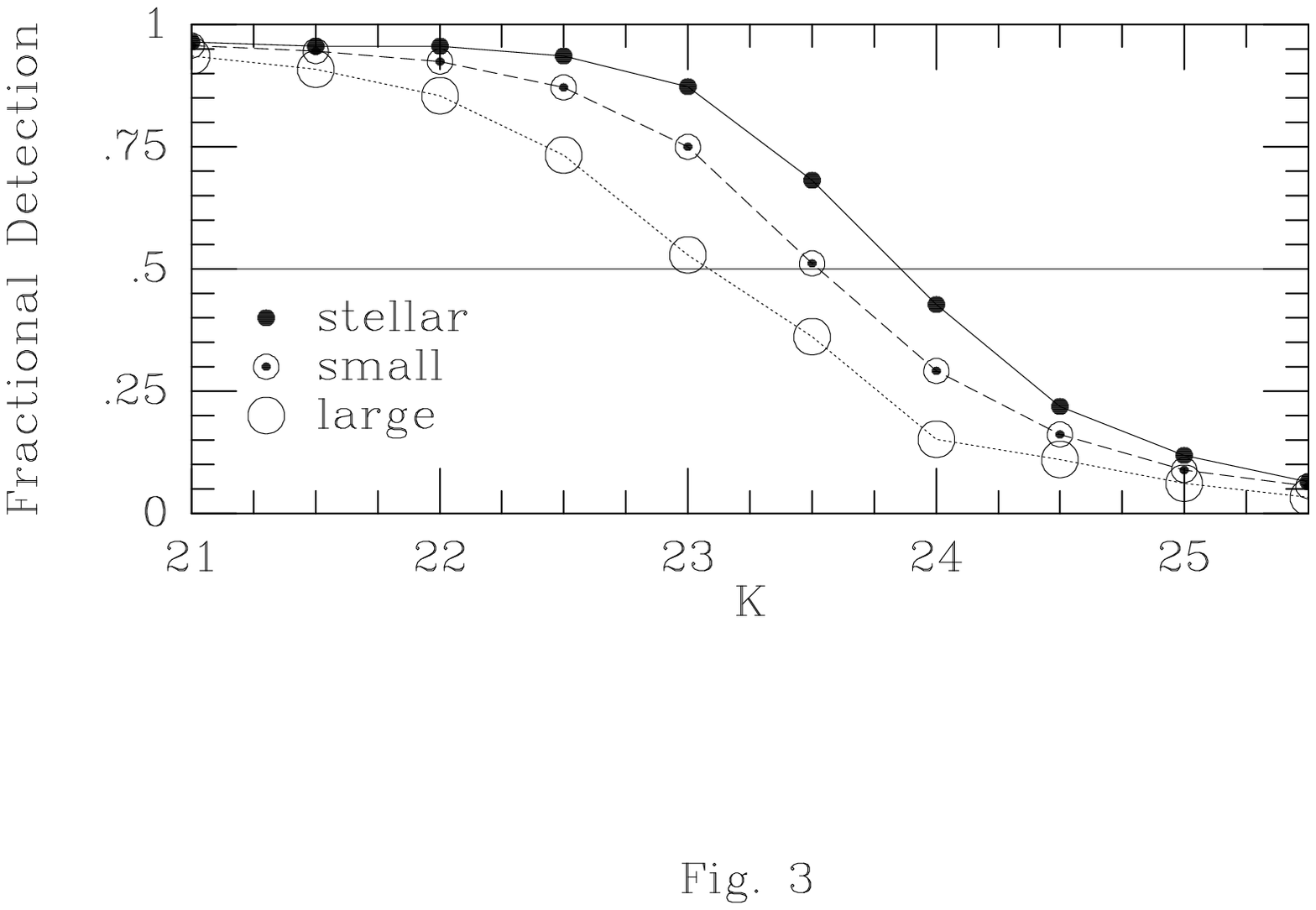}{8in}{0}{100}{100}{-300}{-100}
\end{figure}

\begin{figure}
\plotfiddle{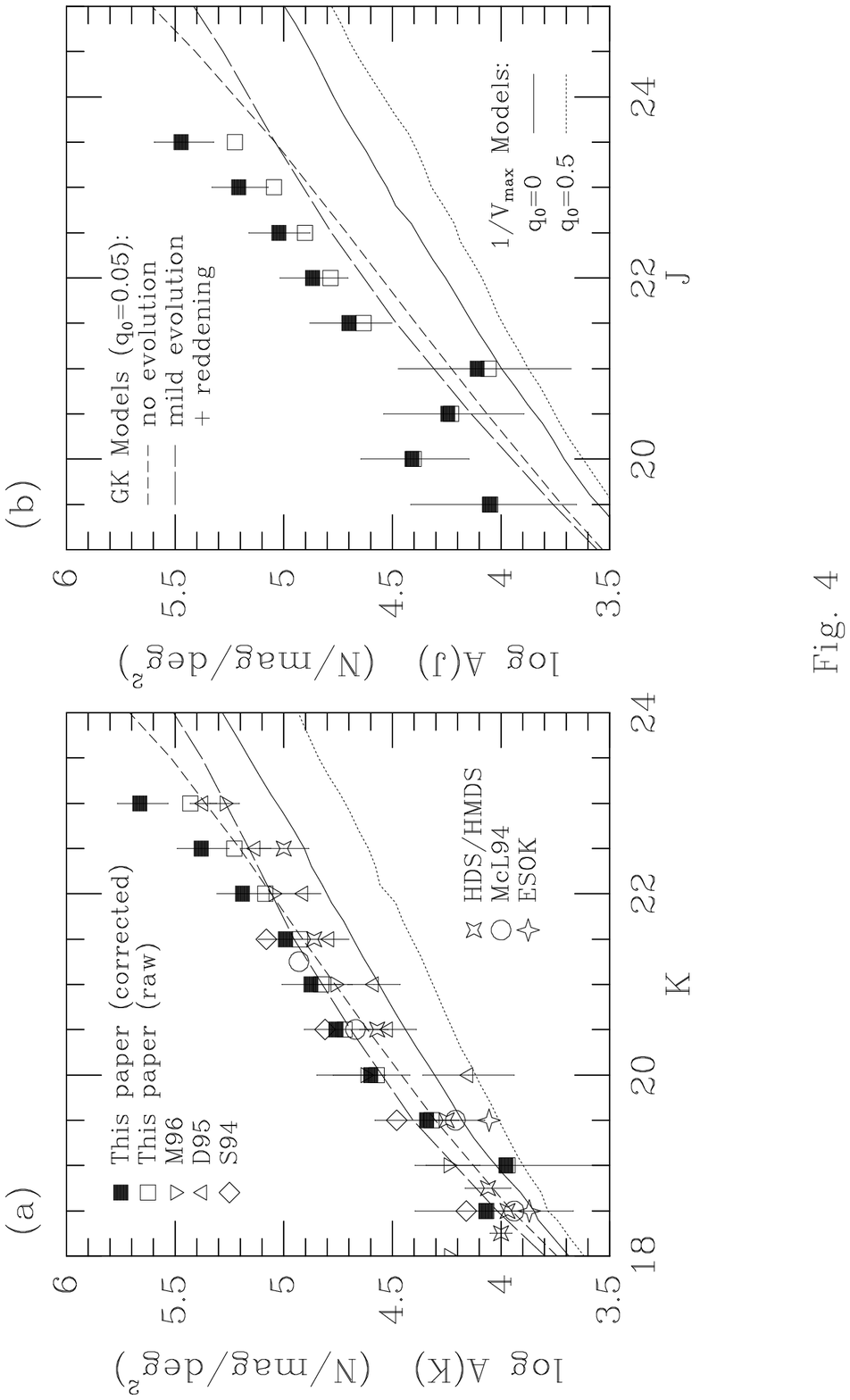}{8in}{0}{100}{100}{-300}{-100}
\end{figure}

\begin{figure}
\plotfiddle{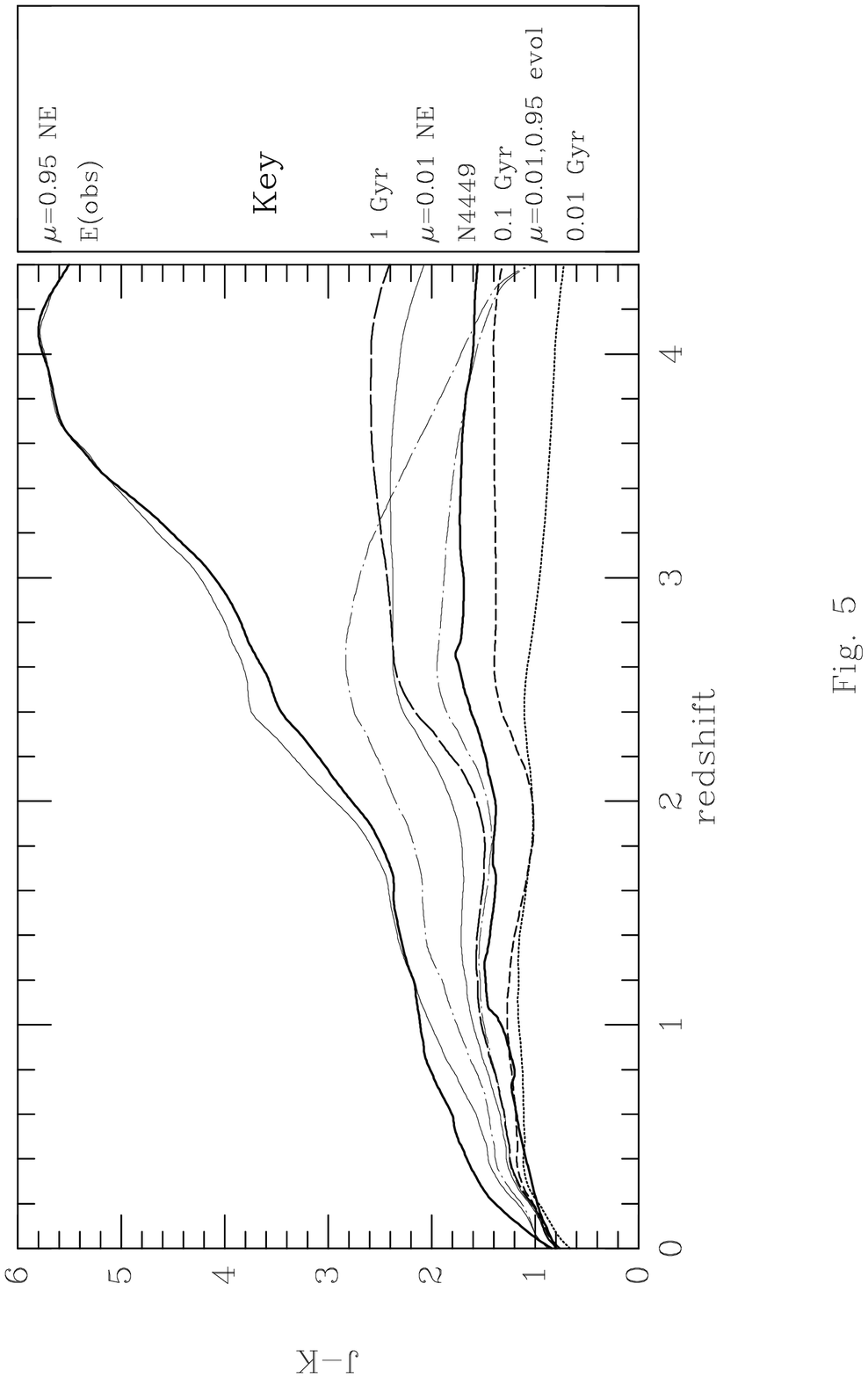}{8in}{0}{100}{100}{-300}{-100}
\end{figure}

\begin{figure}
\plotfiddle{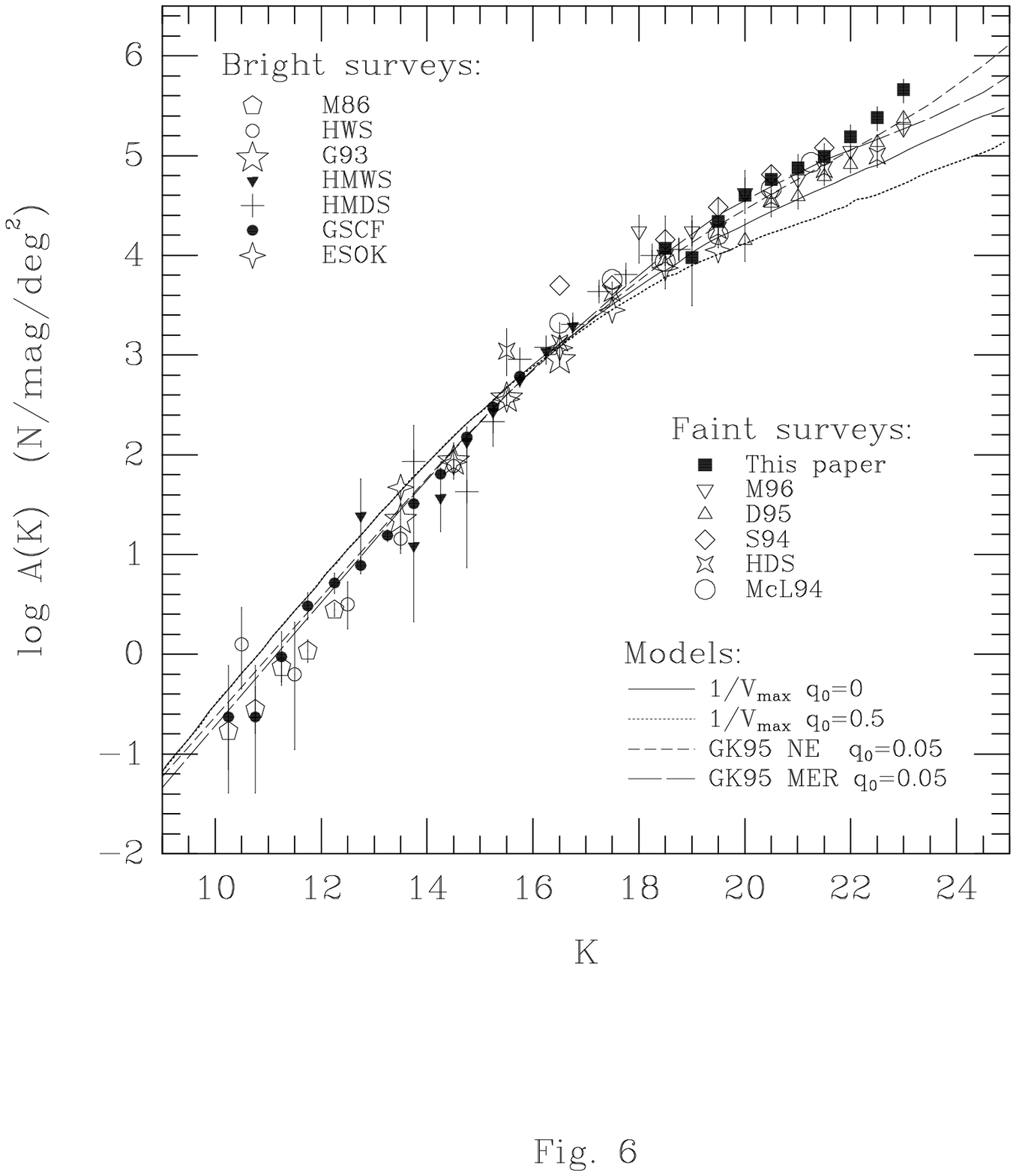}{8in}{0}{100}{100}{-300}{-100}
\end{figure}

\begin{figure}
\plotfiddle{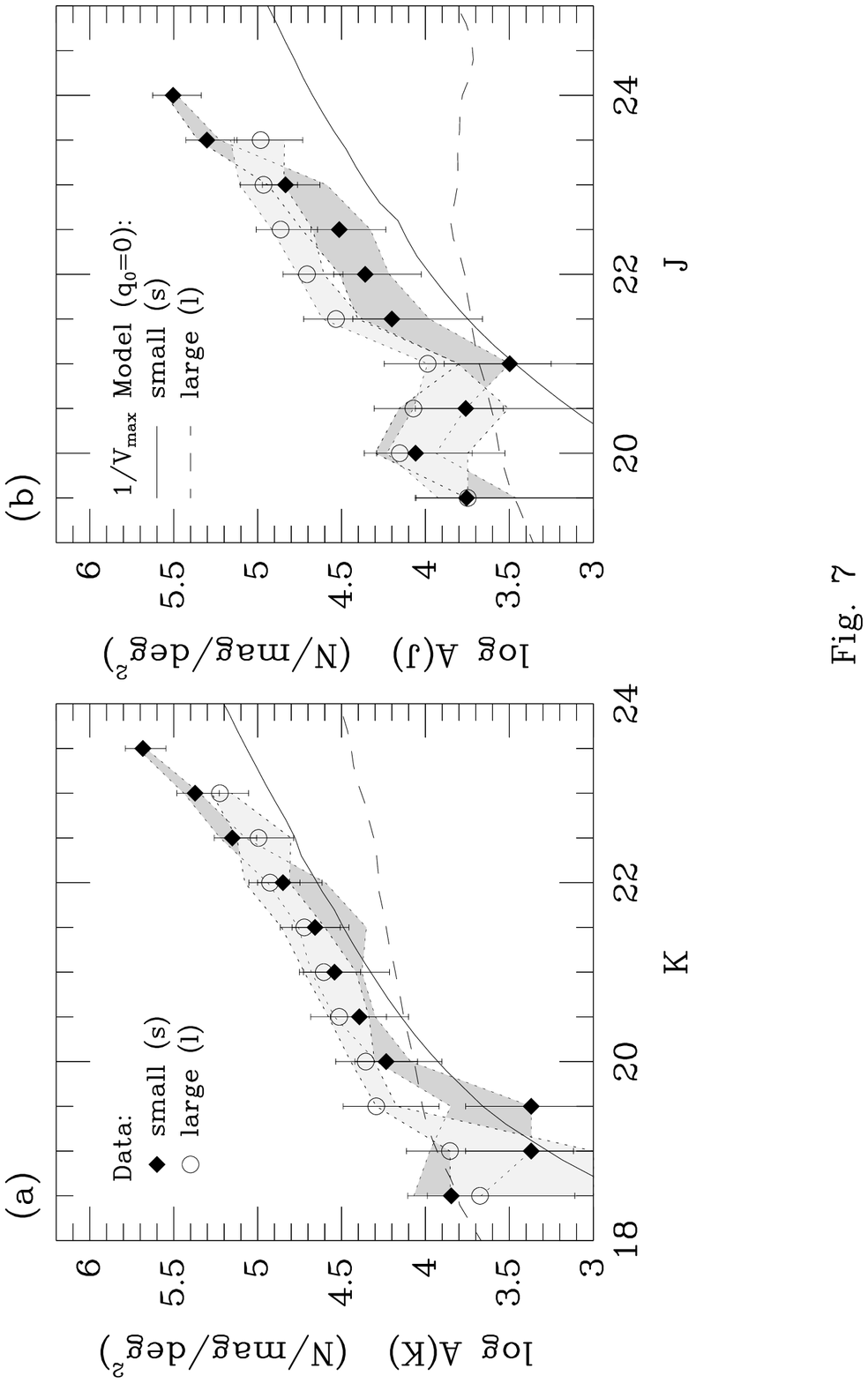}{8in}{0}{100}{100}{-300}{-100}
\end{figure}

\begin{figure}
\plotfiddle{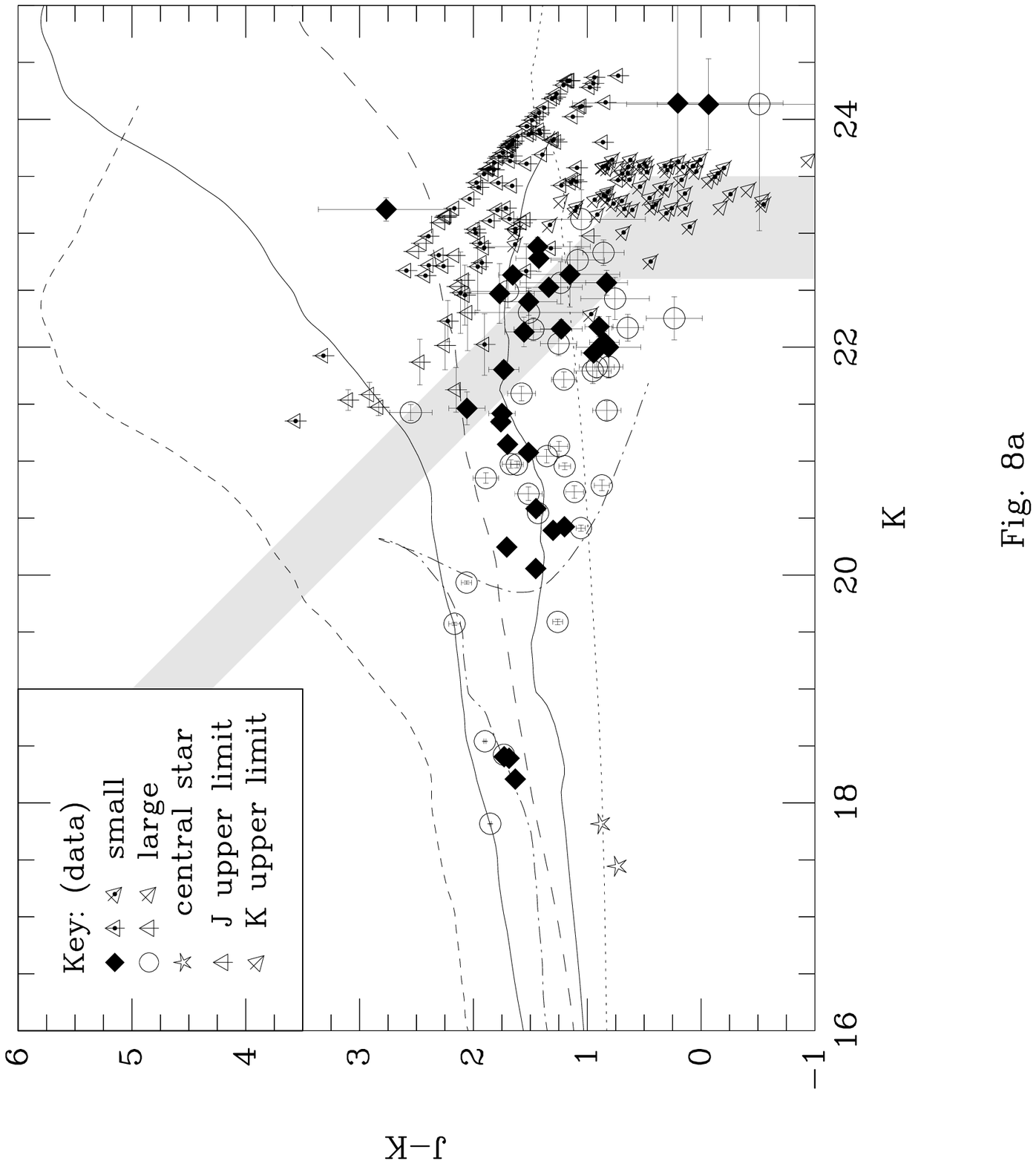}{8in}{0}{100}{100}{-300}{-100}
\end{figure}

\begin{figure}
\plotfiddle{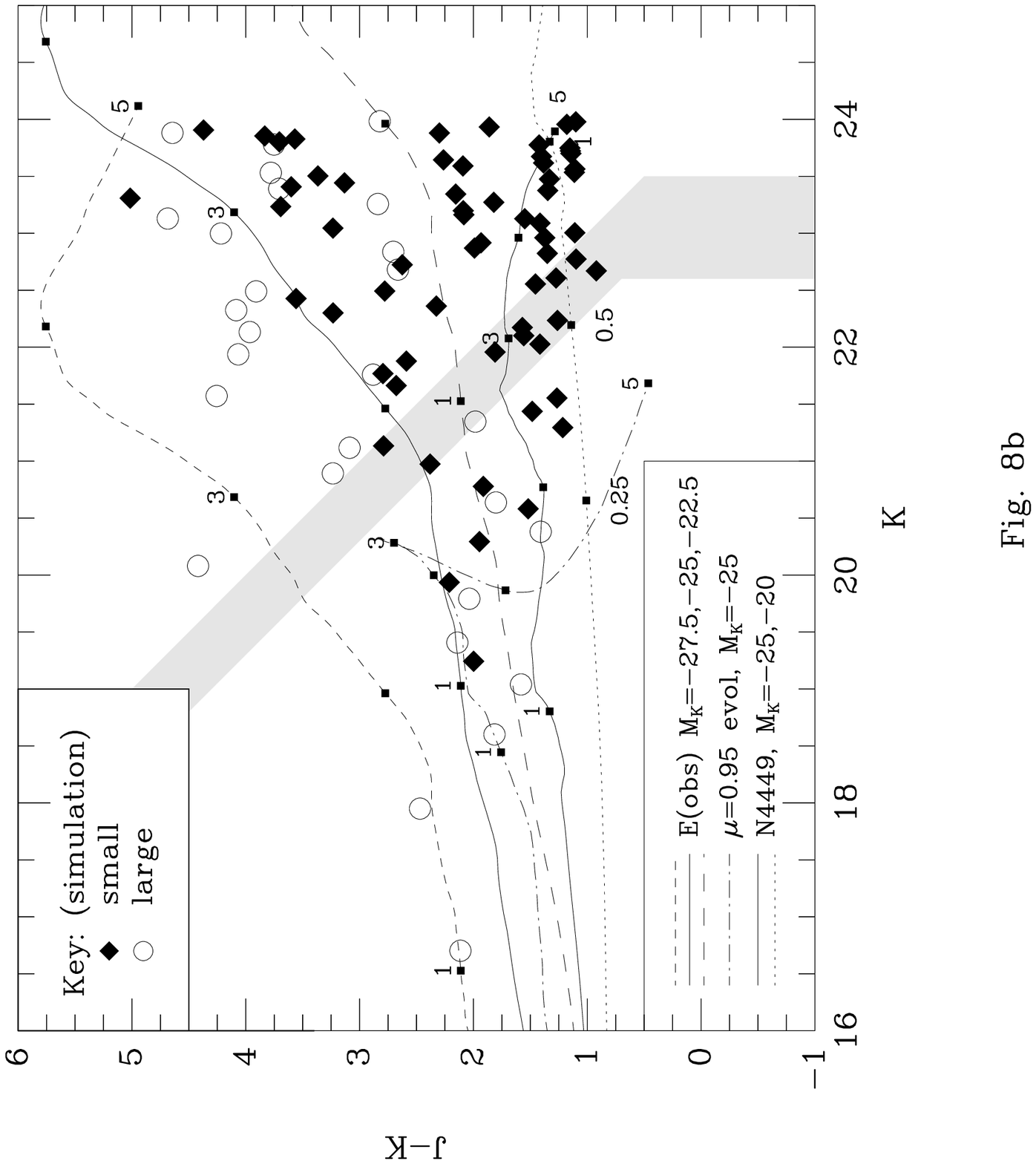}{8in}{0}{100}{100}{-300}{-100}
\end{figure}

\begin{figure}
\plotfiddle{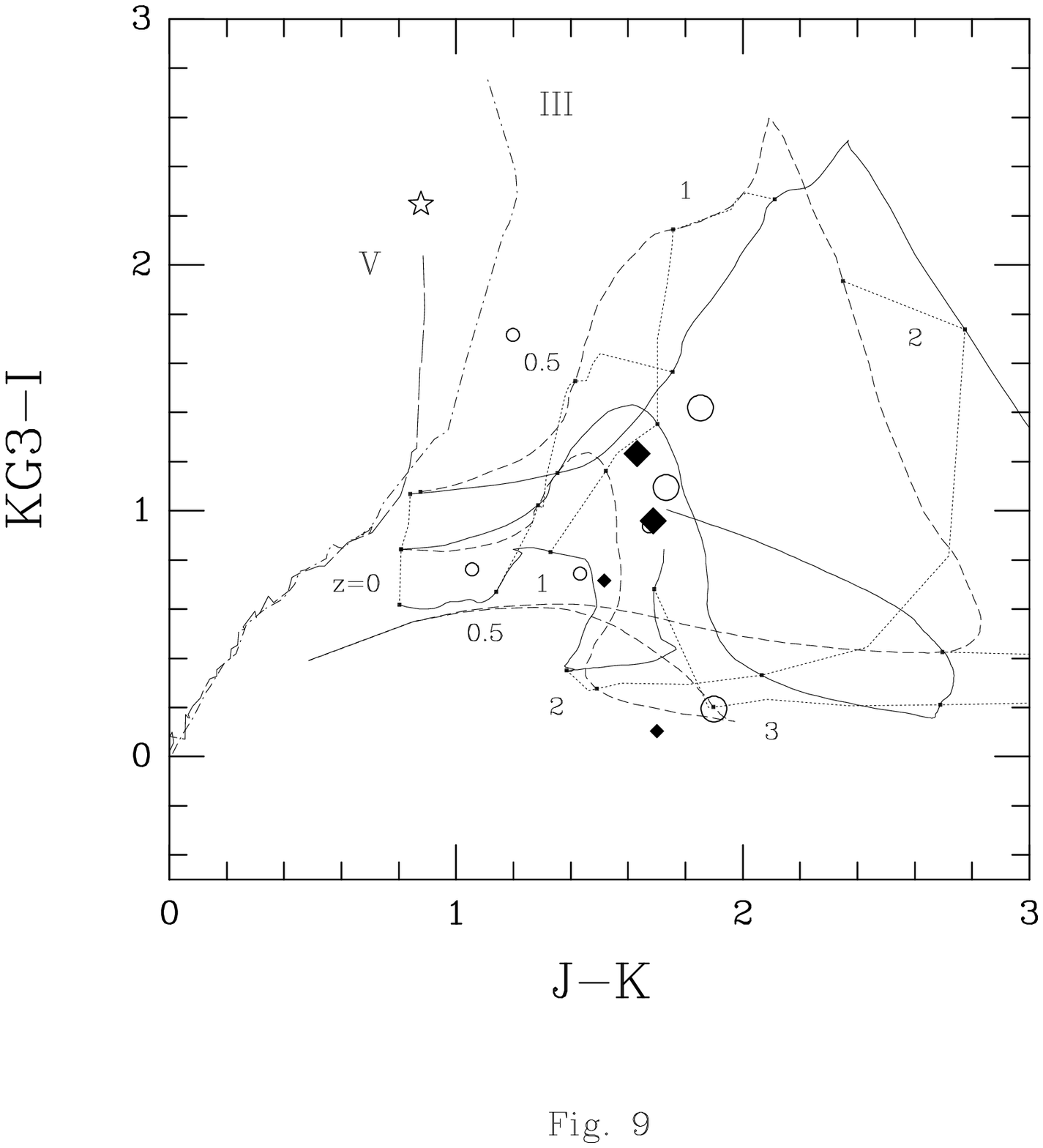}{8in}{0}{100}{100}{-300}{-100}
\end{figure}

\end{document}